\documentclass[reprint,amsmath,amssymb,nofootinbib]{revtex4-2}

\usepackage{graphicx}
\usepackage{bm}
\usepackage{soul} 

\usepackage{braket}

\usepackage{tikz}
\usetikzlibrary{quantikz}
\usepackage{mathtools}
\usepackage[colorlinks=true,citecolor=blue,linkcolor=magenta]{hyperref}

\usepackage{amsmath}
\usepackage{amsthm}
\newtheorem{theorem*}{Theorem}

\newtheorem{lemma*}[theorem*]{Lemma}

\theoremstyle{definition}

\theoremstyle{remark}

\usepackage{algpseudocode}

\newcommand{\RR}{\ensuremath{\mathbb{R}}}

\newcommand{\norm}[1]{\left\lVert#1\right\rVert}
\newcommand{\abs}[1]{\left|#1\right|}

\newcommand{\qhp}[0]{QHP}

\newcommand{\ketbra}[2]{\ensuremath{\left|#1\right\rangle\!\!\left\langle#2\right|}}
\newcommand{\id}{\mathbb{I}}

\renewcommand{\vec}[1]{\boldsymbol{#1}}

\makeatletter 
\renewcommand\onecolumngrid{
\do@columngrid{one}{\@ne}
\def\set@footnotewidth{\onecolumngrid}
\def\footnoterule{\kern-6pt\hrule width 1.5in\kern6pt}
}
\renewcommand\twocolumngrid{
        \def\footnoterule{
        \dimen@\skip\footins\divide\dimen@\thr@@
        \kern-\dimen@\hrule width.5in\kern\dimen@}
        \do@columngrid{mlt}{\tw@}
}
\makeatother

\DeclareMathOperator{\Tr}{Tr}

\newcommand{\zo}[1]{{\color[RGB]{0,0,0}{#1}}}

\begin{document}


\title{On nonlinear transformations in quantum computation}

\author{Zoë Holmes}
\email{zoe.holmes@epfl.ch}
\author{Nolan J. Coble}
\author{Andrew T. Sornborger}
\author{Yi\u{g}it Suba\c{s}\i}
\email{ysubasi@lanl.gov}
\affiliation{Computer, Computational, and Statistical Sciences Division, Los Alamos National Laboratory, Los Alamos, NM 87545, USA.}

\date{\today}

\begin{abstract}
While quantum computers are naturally well-suited to implementing linear operations, it is less clear how to implement nonlinear operations on quantum computers. However, nonlinear subroutines may prove key to a range of applications of quantum computing from solving nonlinear equations to data processing and quantum machine learning. Here we develop \zo{a series of basic subroutines} for 
implementing nonlinear transformations
of input quantum states. Our algorithms are framed around the concept of a \textit{weighted state}, a mathematical entity describing the output of an operational procedure involving both quantum circuits and classical post-processing.
\end{abstract}

\maketitle

\section{\label{sec:level1} Introduction}

Quantum computers are naturally adept at performing linear operations because quantum mechanics is inherently linear. That is, the time evolution of a quantum system is governed by the Schr\"{o}dinger equation, a linear equation. 
Or, equivalently, quantum states evolve under unitary operations which are necessarily linear. However, to exploit the full potential of quantum computing, we need to be able to also twist the arm of quantum devices into implementing nonlinear operations. 

Nonlinear subroutines are likely to play a key role in a range of quantum algorithms. \zo{For example, the ability to efficiently implement nonlinear operations would open up new methods for solving nonlinear equations on quantum hardware~\cite{liu2021efficient,lloyd2020quantum,lin2022koopman}, with applications in areas from fluid dynamics to finance. Alternatively, nonlinear subroutines could prove valuable for developing new techniques for error mitigation by providing a means of amplifying a signal in the presence of background noise~\cite{huggins2020virtual}. Finally, there is much excitement currently about the potential of quantum neural networks and quantum kernel methods~\cite{schuld2021supervised,huang2021power,kubler2021inductive, jerbi2021quantum}}. However, classical neural networks inherit much of their power from the use of nonlinear activation functions. Similarly, kernel methods rely on non-linear encodings. Replicating this on quantum hardware necessitates the ability to implement nonlinear quantum operations.

While quantum mechanics is fundamentally linear, quantum systems often appear to evolve nonlinearly. These apparent nonlinearities are typically induced through measurements and coarsegraining. In the context of quantum computing, in addition to these tools, nonlinear effects can also be introduced using classical post-processing and by collectively manipulating multiple copies of a given input state.

There is a growing body of research into developing new methods for introducing nonlinearities into quantum algorithms. In the context of quantum machine learning, convolutional~\cite{CongQuantum2019} or dissipative~\cite{BeerTraining2020} quantum neural networks, that disregard qubits as the network grows, have been proposed. While these methods do introduce nonlinearities, the exact form of the  nonlinearity is not readily controllable. Algorithms for computing specific nonlinear functions of the elements of a quantum state, have been proposed in a wide range of contexts, including \zo{computing Renyi entropies~\cite{Ekert2002Direct, horodecki2003limits, brun2004measuring}, the negativity of the partial transpose~\cite{carteret2005noiseless, carteret2016estimating, elben2020mixed} and properties of density matrix exponentials~\cite{lloyd2014quantum},} as well as for methods for solving nonlinear equations~\cite{LubaschVariational2020} and studying chaotic systems~\cite{Georgeot2001Exponential, Georgeot2001Stable}. In this paper we tackle the more general problem of 
implementing nonlinear transformations of quantum states. \zo{That is, rather than computing $\Tr[f(\rho)]$ (or $\Tr[f(\rho) M]$) for some function $f$ (and specific measurement $M$), here we focus on implementing $\rho \rightarrow f(\rho)$.} 
This problem has been previously explored using block encoding methodologies~\cite{gilyen2019quantum, dodin2021quantum, guo2021nonlinear}.

In this paper we introduce a new approach for 
preparing nonlinear functions of quantum states, using what we call \textit{weighted states}. These are matrices, describing the output of a quantum instrument, that act like density operators but need not be normalized nor Hermitian. The notion of weighted states is a generalization that includes conditional and marginal states as special cases and shares similarities with virtual state distillation~\cite{huggins2020virtual} and state broadcasting~\cite{Barnum1996Noncommuting, Dariano2005Superbroadcasting, Piani2008No} in the sense that these entities 
allow one to reconstruct expectation values without explicitly preparing the state in a conventional sense. We show that through an appropriate choice of quantum instrument and its inputs it is possible to construct weighted states corresponding to a nonlinear transformation of a density matrix. In particular, our algorithms may be used to implement arbitrary polynomials of the amplitudes of a set of pure input states. 

The outline of the paper is as follows. In Subsection~\ref{sec:Problem} we formally define the functionals of quantum states that we aim to implement and in Subsection~\ref{sec:Framework} we define the notion of weighted states. In Subsection~\ref{sec:QHP} we present an algorithm to implement the Hadamard product of states, in Subsection~\ref{sec:GQT} we present an algorithm to implement the generalized transpose of an input density operator, and in Subsection~\ref{sec:dmp} we describe an algorithm that can be used to prepare a linear combination of 
products of density matrices. We use these algorithms to show how one can implement more general polynomial functions of input quantum states. In Section~\ref{sec:sampling} we analyze the sampling complexity associated with the weighted state approach. We conclude in Section~\ref{sec:discussion}.

\section{Weighted States}
\label{sec:ES}

\begin{figure*}
\centering
\includegraphics[]{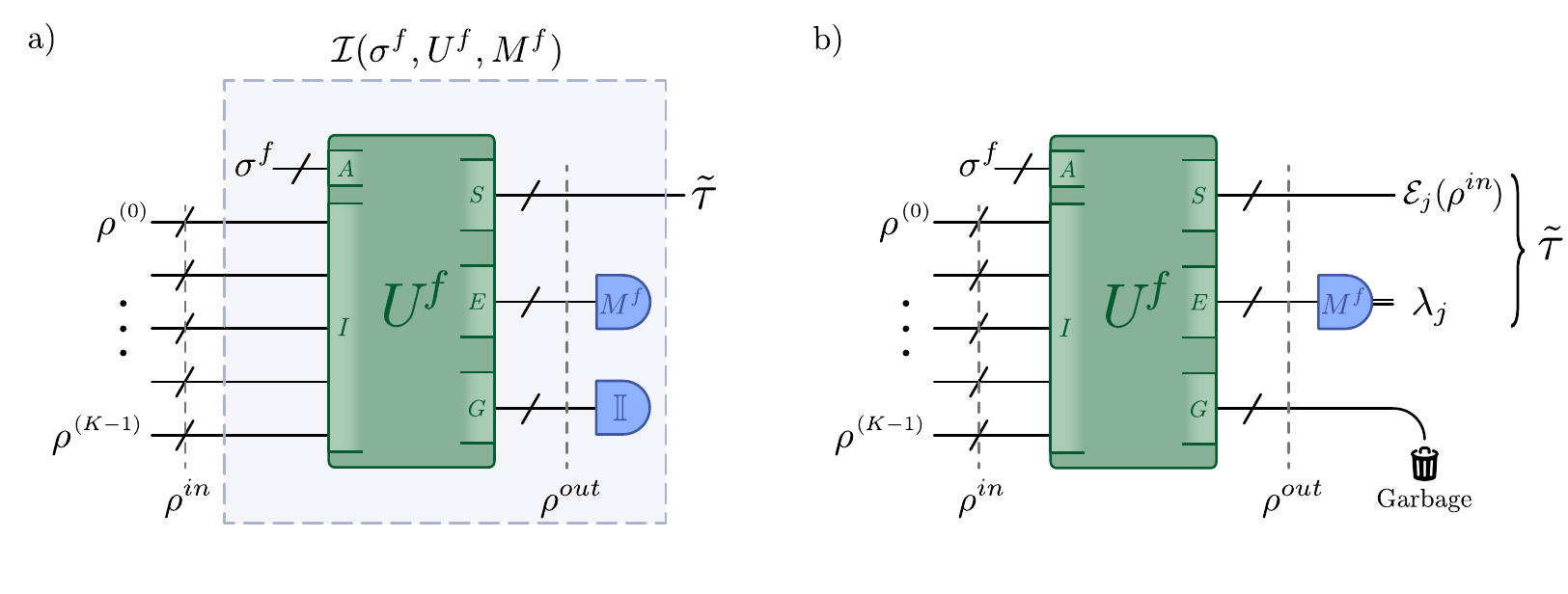}
\caption{\textbf{Weighted state framework.} Here we illustrate two equivalent ways of interpreting weighted states. a) A weighted state $\tilde{\tau}=\Tr_{EG}(\rho^\text{out} (\mathbb{I}_S \otimes M^f_E \otimes\mathbb{I}_G) )$ is equivalent to the output of a composite state $\rho^\text{out}$ after the measurement $M^f\otimes \mathbb{I}$ on the environment and garbage auxiliary systems. b) A weighted state $\tilde{\tau} = \sum_j \lambda_j \mathcal{E}_j(\rho^\text{in})$ is the output of a quantum instrument where conditional on measuring the $j$'th output on the environment register $E$, the quantum output in register $S$ is weighted by $\lambda_j$.}
\label{fig:es}
\end{figure*}

\subsection{Statement of problem}\label{sec:Problem}

Given a set of density operators as input
$\rho^\text{in} = \rho^{(0)} \otimes \dots \otimes \rho^{(K-1)}$, we are interested in preparing weighted states (to be explained in detail later) of the form
\begin{align}
    \label{eq:es_def}
    \rho^\text{in} &\rightarrow \tilde{\tau} = \sum_{i,j=0}^{d_\text{out}-1} f(\vec{v}_{\rho^\text{in}})_{i,j} \ketbra{i}{j}\, ,
\end{align}
where $\vec{v}_{\rho^\text{in}}$ is the vector containing all entries of the input density operators in the computational basis, i.e.  $\vec{v}_{\rho^\text{in}} = ( \rho^{(0)}_{0,0}, \rho^{(0)}_{0,1}, \, ... , \rho^{(K-1)}_{d,d} )$ with $\rho^{(k)}_{i,j}=\braket{i|\rho^{(k)}|j}$, and $f: \mathbb{C}^{k d^2} \rightarrow \mathbb{C}^{d_\text{out}^2}$ is a multilinear function of its arguments.
Here $d$ is the Hilbert space dimension\footnote{The general framework allows for input systems with different Hilbert space dimensions, however we won't be considering such cases in this work.} of the individual inputs $\rho^{(k)}$ and $d_\text{out}$ is the Hilbert space dimensions of the output system $\tilde{\tau}$.

Crucially, if $\rho^{\rm in}$ includes multiple copies of the same state $\rho$, and we consider $f$ to be a function of the \textit{unique} input states, then the transformation functions will in general be nonlinear in the elements of $\rho$. For example, in the case where the input states are all identical, i.e. $\rho_{\rm in} = \rho^{\otimes K}$, we can consider the transformation 
\begin{align}
    \label{eq:es_nonlin_def}
    \rho &\rightarrow \tilde{\tau} = \sum_{i,j=0}^{d _\text{out}-1} g(\vec{u}_{\rho})_{i,j} \ketbra{i}{j}\, ,
\end{align}
where $\vec{u}_{\rho} = ( \rho_{0,0}, \rho_{0,1}, \, ... , \rho_{d,d} )$ is now the vector containing all entries of $\rho$ in the computational basis,
and the transformation functions $g: \mathbb{C}^{d^2} \rightarrow \mathbb{C}^{d_\text{out}^2}$ are in general nonlinear. \zo{In this manuscript, we develop a series of basic subroutines that can be concatenated to implement (a subset of) non-linear operations of this form.}

Note that this operation may depend on the chosen computational basis. In general, $\tilde{\tau}$ is not normalized, positive definite, and might not even be Hermitian. We use tilde to indicate such quantum objects that are not proper quantum states that exist on a register of a quantum computer.

A special case is pure states. In this case we assume as input a set of pure states
$\ket{\psi^\text{in}}=\ket{\psi^{(0)}}\otimes \dots \otimes \ket{\psi^{(K-1)}}$, and we are interested in transformations of the form
\begin{align}
\label{eq:es_def_pure}
\ket{\psi^\text{in}} &\rightarrow \ket{\tilde{\phi}} = \sum_i h(\vec{v}_{\psi^\text{in}})_i \ket{i}\, ,
\end{align}
where $\vec{v}_{\psi^\text{in}}$ is the vector containing all amplitudes of the input states in the computational basis, i.e.  $\vec{v}_{\psi^\text{in}} = ( \psi^{(0)}_{0}, \psi^{(0)}_{1}, \, ... , \psi^{(K-1)}_{d_{\rm in}} )$ where $\psi^{(k)}_{i}=\braket{i|\psi^{(k)}}$, and $h: \mathbb{C}^{k d} \rightarrow \mathbb{C}^{d_\text{out}}$ is a multilinear function of its arguments. Again, if $|\psi^{\rm in} \rangle$ includes multiple copies of the same state $|\psi \rangle$, $h$ will in general be nonlinear with respect to the amplitudes of $| \psi \rangle$. 

\subsection{General framework}\label{sec:Framework}

In this section, we introduce the concept of `weighted states' which provides the backbone of all the algorithms presented later on. To motivate this concept we start by emphasizing that one only ever has access to quantum states through measurement outcomes. Thus, suppose we are interested in the outcome of measurements on the state $\tau$, it is not strictly necessary to prepare $\tau$, rather it suffices to propose an operational strategy for calculating $\langle O \rangle_\tau = \Tr[\tau O]$ for any measurement operator $O$.

When the state we are interested in is a nonlinear functional of the input state as in Eq.~\eqref{eq:es_nonlin_def}, we cannot strictly prepare it unless it happens to be a proper quantum state, i.e. unless $\tilde{\tau} =\tau$ is positive semi-definite, Hermitian matrix with unit trace. 
Instead, we first prepare a quantum state in an extended Hilbert space composed of Ancilla register ($A$) and an Input register ($I$). As shown in Fig.~\ref{fig:es}a), the initial composite state of the registers $\sigma^f_A\otimes \rho^\text{in}_I$ evolves unitarily under $U^f$ such that the final state is  $\rho^\text{out}=U^f (\sigma^f_A\otimes \rho^\text{in}_I)U^{f\dagger}$. Here $\sigma^f$ is the ancilla state to be chosen as part of the algorithm. We then regroup the registers of $\rho^\text{out}$ into a system $(S)$, an environment $(E)$, and the garbage $(G)$. Finally, we pick a measurement operator $M^f$ on the environment such that the following is satisfied:
\begin{align}
    \label{eq:map}
    \Tr[\tilde{\tau} O] &= \Tr_{SEG}[\rho^\text{out} (O \otimes M^f_E\otimes \mathbb{I}_G)]   \ \ \ \forall \ \ O \, .
\end{align}
In other words, we map the expectation value of any system operator $O$ in the (possibly unphysical) state $\tilde{\tau}$ to the expectation value of the operator $O \otimes M^f\otimes \mathbb{I}$ in the composite state $\rho^\text{out}$. 
Since Eq.~\eqref{eq:map} is true for any system operator $O$, we can also express the weighted state as
\begin{equation}
    \label{eq:es1}
    \tilde{\tau} = \Tr_{EG}(\rho^\text{out} (\mathbb{I}_S \otimes M^f_E \otimes\mathbb{I}_G) ) \, .
\end{equation}
This could be interpreted as a generalization of the partial trace map.
We note that the measurement of $O$ need not be made straightaway, rather the weighted state $\tilde{\tau}$ can be used as the input to another algorithm and so may be further processed by additional subroutines before measurement.

Alternatively, one can think of weighted states in terms of quantum instruments which are quantum operations with both classical and quantum outputs~\cite{davies1970Operational}. Any quantum instrument can be realized by applying a joint unitary transformation on the input system plus an ancilla, followed by a projective measurement on part of the resulting joint system as in Fig.~\ref{fig:es}b). We indicate such an instrument\footnote{Clearly, there is a degeneracy in the choice of the set \{$\sigma^f,U^f$,$M^f$\} and different choices can result in the same instrument. One could eliminate this degeneracy by adopting the convention that $M^f$ is diagonal and $\sigma^f=\ketbra{0}{0}$, which is always possible as the unitary diagonalizing $M^f$ and the one that prepares the purification of $\sigma^f$ can always be absorbed into $U^f$. However, we believe that this convention makes it harder to build intuition about the algorithms we describe later in the paper.} as $\mathcal{I}(\sigma^f,U^f,M^f)$.  
In general, the classical and quantum outputs are correlated. 
If the $j$'th possible measurement outcome $\lambda_j$ is observed in a realization of the measurement of $M^f$, the conditional quantum state on the $S$ register of the joint system is given by $\mathcal{E}_j(\rho^\text{in})/\Tr(\mathcal{E}_j(\rho^\text{in}))$, where $\mathcal{E}_j(\square_I)=\Tr_{EG}(U^f(\sigma^f_A\otimes \square_I) U^{f\dagger} (\mathbb{I}_S\otimes \ketbra{\lambda_j}{\lambda_j}_E\otimes\mathbb{I}_G))$ is a completely positive trace-non-increasing map and $\Tr_S(\mathcal{E}_j(\rho^\text{in}))$ is the probability $p_j$ of observing said outcome~\cite{oreshkov2012Quantum}. Since probabilities add up to one, $\mathcal{E} = \sum_j \mathcal{E}_j$ is completely positive and trace preserving and $\mathcal{E}(\rho^\text{in})=\Tr_{EG}(\rho^\text{out})$ describes the marginal state on the $S$ register. What we refer to as the weighted state can be expressed as
\begin{align}
    \label{eq:es2}
    \tilde{\tau} &= \sum_j \lambda_j p_j \frac{\mathcal{E}_j(\rho^\text{in})}{\Tr(\mathcal{E}_j(\rho^\text{in}))} =  \sum_j \lambda_j \mathcal{E}_j(\rho^\text{in}) \; .
\end{align}
In this manner, the weighted state is obtained by weighting the quantum outcomes of the quantum instrument with its  classical outcomes. In practice this weighting is implemented in post-processing.
It is evident that conditional and marginal states are special cases of weighted states whereby the measurement operator $M_f$ is a projector and identity, respectively. 
In many ways, one can think of the use of conditional states as being the analog of rejection sampling whereas weighted states are analogous to importance sampling.

The goal is then to find a quantum instrument $\mathcal{I}(\sigma^f,U^f,M^f)$ that
implements the transformation of Eq.~\eqref{eq:es_def} for a given function $f$ for all inputs $\rho^\text{in}$. 
First, let us assume that $M^f$ is a normal operator and hence can be diagonalized by a unitary. A subsequent measurement in the computational basis, with outcome $j$, is interpreted as the measurement of $\lambda_j$, the $j$'th eigenvalue of $M^f$. 
The physical states $\mathcal{E}_j(\rho^\text{in})/\Tr(\mathcal{E}_j(\rho^\text{in}))$ can then be used, together with the classical outputs $\lambda_j$ of the instrument, to emulate any quantum computation that involves the (possibly unphysical) state $\tilde{\tau}$. 

The method described in the previous paragraph assumes $M^f$ is a normal operator. More generally, one can emulate an arbitrary operator $M^f$ by randomizing over quantum instruments. Let $M^{f,(k)}$ be a normal operator corresponding to the $k$'th instrument and $q_k$ be the probability with which this instrument is sampled. Then the net effect of using a randomized instrument is captured by $M^f=\sum_k q_k M^{f,(k)}$ in Eqs.~(\ref{eq:map},\ref{eq:es1}). Note that, in principle, any operator can be represented this way as $M^f=(1/2)(M^f+M^{f\dagger})+(1/2)(M^f-M^{f\dagger})$, where both the Hermitian and skew-Hermitian parts of $M$ are normal (but other decompositions may prove to be more efficient). 
In Appendix.~\ref{ap:beyondnormal} we show that this randomization is not necessary and in fact a single quantum instrument can always be constructed that implements the transformation that a fictitious instrument with an arbitrary non-normal $M^f$ would. Hence, unless stated otherwise, we will henceforth assume $M^f$ is normal. 

Finally, it is important to note that $\rho^\text{in}$ can itself be a weighted state and Eqs.~(\ref{eq:es1},\ref{eq:es2}) still hold in this case. Thus algorithms implementing basic nonlinear transformations can be concatenated to prepare a large class of weighted states of the form Eqs.~(\ref{eq:es_def}, \ref{eq:es_def_pure}) corresponding to complex nonlinear transformations of input states. 
Below we will describe three such basic algorithms. The first algorithm multiplies two input state density operators entry-wise in the computational basis (Hadamard product), the second implements a generalization of the transpose operation, and the third algorithm outputs polynomials of input density matrices.  

In the rest of the paper we will drop the tilde on $\tilde{\tau}$ as there is no need to make a distinction between physical and weighted states in the framework. 
We will also drop the superscript $f$ and let \{$\sigma^f,U^f,M^f$\}$\rightarrow$
\{$\sigma,U,M$\} for brevity. It is always to be understood that this set depends on $f$. 

\subsection{Quantum Hadamard Product (QHP)}\label{sec:QHP}

We define the Quantum Hadamard Product (QHP) of two states  $\rho^{(0)}$ and $\rho^{(1)}$ as
\begin{align}
    \label{eq:QHP}
    \rho^{(0)} \odot \rho^{(1)} = \sum_{ij} \rho^{(0)}_{ij} \rho^{(1)}_{ij} \ketbra{i}{j} \, .
\end{align}
That is, the matrix elements of the QHP in the computational basis are the product of the matrix elements of the input density operators.

\begin{figure}
\centering
\includegraphics[]{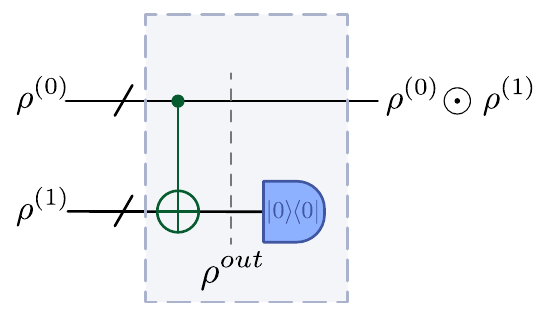}
\caption{\textbf{Quantum Hadamard Product (\qhp ).} Here we show the depth 1 circuit to implement \qhp. Note that the CNOT gates act on all pairs of qubits.}
\label{fig:qhp}
\end{figure}

In Fig.~\ref{fig:qhp} we present a circuit for implementing the Hadamard product.
The output state after applying the ladder of CNOT gates, $U = \text{CNOT}^{\otimes n}$, is 
\begin{align}
    \rho^\text{out} 
    &= \sum_{i_0,j_0,i_{1},j_{1}}\rho^{(0)}_{i_0j_0}\rho^{(1)}_{i_{1}j_{1}}\ket{i_0,i_0\oplus i_{1}}\!\bra{j_0,j_0\oplus j_{1}} 
\end{align}
The measurement operator is $M=| 0 \rangle \langle 0 |$ and there is no garbage register. It is straightforward to verify that the weighted state is the desired Quantum Hadamard Product, that is we have
\begin{align}
\label{eq:QHP}
    \tau = \Tr_{EG}(\rho^\text{out} (\mathbb{I}_S \otimes M ) ) = \rho^{(0)} \odot \rho^{(1)} \, .
\end{align}
We note that variants of this circuit for preparing normalized Hadamard product states have been previously been proposed in Refs.~\cite{BechmannPasquinucci1998NonLinear, Ludovico2016Bipartite}.

Note that $M$ is a projector. As such it amounts to performing a post-selection. In that viewpoint, one would obtain the normalized state $\rho^{(0)} \odot \rho^{(1)}/\Tr[\rho^{(0)} \odot \rho^{(1)}]$ whenever the second register is measured in the all 0's state. Then one measures the observable $O$ in this state. This process is repeated until enough statistics is collected. Finally, one would have to multiply the outcome with the normalization factor  $\Tr[\rho^{(0)} \odot \rho^{(1)}]$ which can be estimated by the relative frequency of the success of post-selection. Thus the weighted state formalism includes methods based on post-selection as a special case, whereby $M$ is proportional to a projector. 

\begin{figure*}
\centering
\includegraphics[]{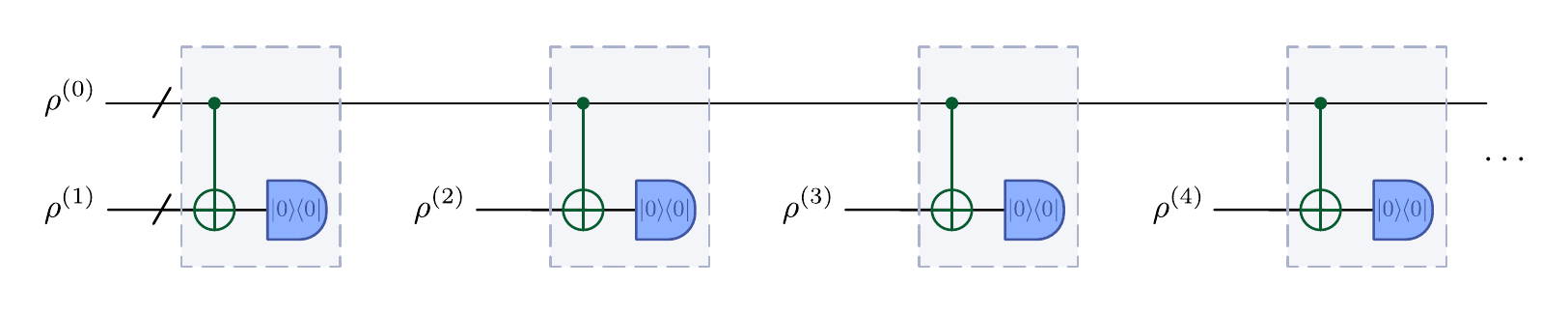}
\caption{\textbf{Iterated Quantum Hadamard Product.} The circuit consists of $n-1$ repetitions of the QHP subroutine shown in Fig.~\ref{fig:qhp}. This circuit can be used with $\rho^{(1)} = \rho^{(2)} = ... = \rho^{(k)} = | \psi \rangle \langle \psi |$ for preparing a quantum state $\sum_i \psi_i^k \ket{i}$ in which all the amplitudes in the computational basis are raised to some power $k$. We note that one could also reset alternating qubits. These two instruments would produce the same final weighted state on an ideal device but may perform differently in the presence of noise~\cite{Yirka2021qubitefficient}.} 
\label{fig:PoS}
\end{figure*}

In the special case where the inputs to QHP are pure states, i.e. $\rho^{(0)}=\ketbra{\psi^{(0)}}{\psi^{(0)}}$ and $\rho^{(1)}=\ketbra{\psi^{(1)}}{\psi^{(1)}}$, the QHP transforms them into another pure state given by
\begin{align}
    \ket{\phi} = \ket{\psi^{(0)}\odot \psi^{(1)}}\equiv \sum_i \psi^{(0)}_i \psi^{(1)}_i \ket{i}\, . 
\end{align}
Weighted states $\ket{\psi^{(0)}\odot\psi^{(1)}\odot \dots\psi^{(K-1)}}$ that are the Hadamard product of $K$ states can be prepared by using the QHP algorithm iteratively as shown in Fig.~\ref{fig:PoS}. If the input states are defined on $n$ qubits, this algorithm can be implemented using $2n$ qubits, independent of $K$, by resetting and reusing qubits~\cite{Yirka2021qubitefficient}. 
Of particular interest is the potential to use iterative applications of QHP to generate powers of a state, that is to prepare
\begin{align}
    \label{eq:nthpower}
    \ket{\underbrace{\psi\odot\psi\odot\dots\odot\psi}_{p \text{ times}}}\equiv\ket{\psi^p} &\equiv \sum_i \psi_i^p \ket{i} \,.
\end{align}
More generally, using the circuit in Fig.~\ref{fig:PoS} with different input states we can prepare weighted states whose amplitudes are products of the amplitudes of different states such as
\begin{align}
    \sum_{i} \prod_j  \left(\psi^{(j)}_i\right)^{p_{j}}  \ket{i} \, .
\end{align}
Such products may be used as building blocks for preparing arbitrary polynomial functions.

If we treat $\rho^{(0)}\rightarrow \sigma$ as part of the ancilla register $(A)$ and only $\rho^{(1)}\rightarrow \rho$ as the input $(I)$ of the quantum instrument, then the same quantum circuit as QHP shown in Fig.~\ref{fig:qhp} implements the following linear transformation
\begin{align}
    \rho \rightarrow \sigma\odot \rho = \sum_{ij} \sigma_{ij} \rho_{ij} \ket{i}\!\bra{j}\, .
\end{align}
The only difference between the two quantum instruments is the interpretation of what constitutes the input of the instrument.

\subsection{Generalized Quantum Transpose (GQT)}\label{sec:GQT}

In this section we present an algorithm for implementing a transformation related to the QHP that we call the Generalized Quantum Transpose (GQT). Given an input state $\rho$ and ancillary state $\sigma$, the GQT is defined as
\begin{align}
    \label{eq:QHP}
    \rho_{\sigma}^{(T)} := \sigma \odot \rho^T  \, .
\end{align}
That is, the output of GQT is the transpose of the input density operator $\rho$ in the computational basis, with the elements of $\rho^T$ weighted by the elements of $\sigma$ on an element-by-element basis. When $\sigma$ is chosen to be the plus state $\sigma = | + \rangle \langle + |$ with $| + \rangle = \frac{1}{\sqrt{d}} \sum_i |i \rangle$, the output is the transpose of $\rho$ up to the dimension of the input, that is, $\rho_{|+\rangle\langle+|}^{(T)} = \frac{1}{d} \rho^{T}$.
In the special case of pure input states, i.e. $\rho=\ketbra{\psi}{\psi}$, the GQT implements a weighted complex conjugation operation such that the output is $\sigma \odot \ketbra{\psi^*}{\psi^*}$. 

\begin{figure}
\centering
\includegraphics[]{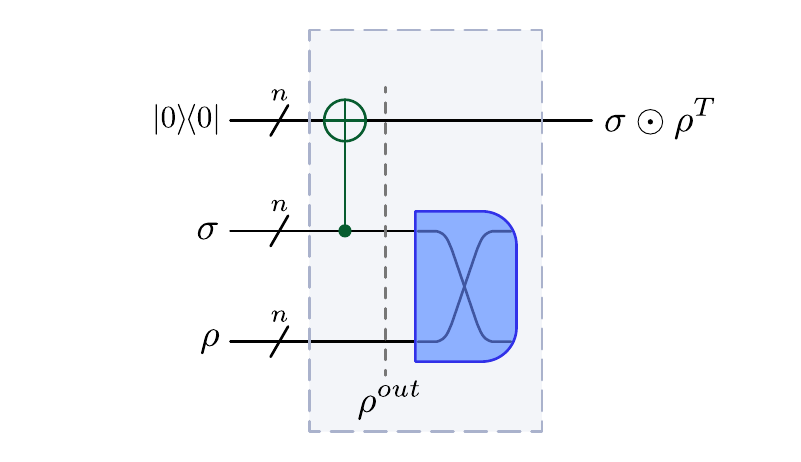}
\caption{\textbf{Generalized Quantum Transpose (GQT).} Here we show the depth 3 circuit to implement GQT. The purple box indicates a SWAP measurement (for breakdown see Fig.~\ref{fig:SWAP}).}
\label{fig:gqt}
\end{figure}

\begin{figure}
\centering
\includegraphics[]{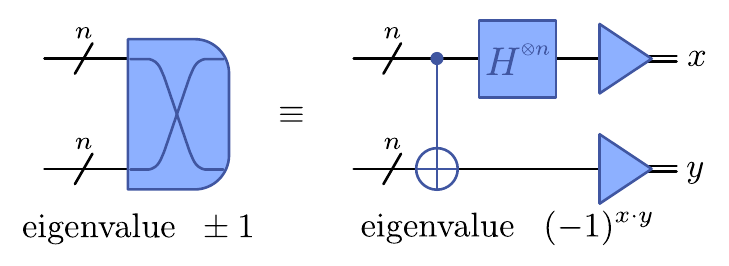}
\caption{\textbf{Swap measurement.} The implementation of the SWAP measurement with a circuit of depth 2~\cite{cincio2018learning}. The readouts are $x=x_1,\dots,x_n$ and $y=y_1,\dots,y_n$, and $x\cdot y=x_1 y_1+\dots+x_n y_n$.}
\label{fig:SWAP}
\end{figure}

In Fig.~\ref{fig:gqt} we present a circuit for implementing the GQT. Specifically, the weighted state prepared by this circuit can be shown to be:
\begin{align}
    \rho^\text{out} &= \sum_{ii'jj'} \sigma_{ij} \rho_{i'j'} \ketbra{iii'}{jjj'} \\
    \tilde{\tau} &= \Tr_{2,3}(\rho^\text{out} (\mathbb{I} \otimes \text{SWAP}) )\\
    &= \sum_{iji'j'} \sigma_{ij} \rho_{i'j'}  \bra{ii'}\text{SWAP}\ket{jj'}  \ket{i}\!\bra{j} \\
    &= \sum_{ij} \sigma_{ij} \rho_{ji}  \ket{i}\!\bra{j} = \sigma \odot {\rho}^{T}\, ,
\end{align}
where the trace in second equation is taken over the last two registers. This circuit may be viewed as a implementing a generalized form of quantum teleportation, as discussed further in Appendix~\ref{app:teleportation}.
Note that the Bell basis states are eigenstates of the SWAP operator with eigenvalues $\pm 1$.
Hence the SWAP measurement on the bottom two registers can be realized with a depth two quantum circuit as seen in Fig.~\ref{fig:SWAP}~\cite{cincio2018learning}. 

GQT can be used in tandem with the other algorithms described in this work to enlarge the set of functions $f$ that can be implemented. Although GQT, in its current formulation, implements a linear transformation, if the ancilla state is chosen to be $\rho$, or GQT is applied iteratively taking multiple copies of the state $\rho$ as inputs, the resulting transformation will be nonlinear in $\rho$. It is worth noting that GQT can be applied to a subsystem of a larger system to implement a partial transpose. Thus GQT could be leveraged to witness entanglement in mixed quantum states~\cite{PeresSeparability1996, HorodeckiSeparability1996}.

\medskip

\subsection{Quantum State Polynomial (QSP)}\label{sec:dmp}

\begin{figure}
\centering
\includegraphics[]{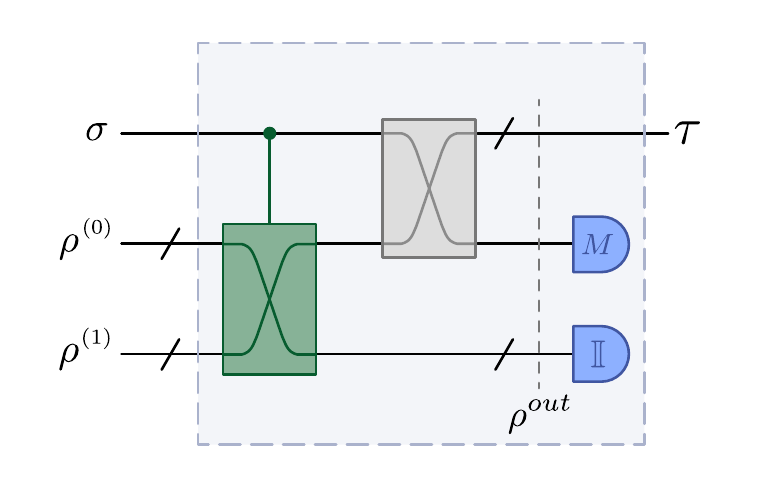}
\caption{\textbf{Quantum State Polynomial (QSP).} Here we show the circuit to implement QSP. The controlled SWAP gate can be implemented with three Toffoli gates. The final swap does not need to be implemented but rather is included to match the register labelling convention chosen in Fig.~\ref{fig:es}. 
}
\label{fig:QSP}
\end{figure}

In this section we present an algorithm that takes as input two states $\rho^{(0)}$ and $\rho^{(1)}$ and prepares weighted states
\begin{align}
    \nonumber
    \tilde{\tau} = &\alpha_{00} \rho^{(0)}+ \alpha_{11} \rho^{(1)} \\
    \label{eq:StatePoly}
    &+ \alpha_{01} \rho^{(0)} \rho^{(1)}+ \alpha_{10} \rho^{(1)} \rho^{(0)} \, .
\end{align}
Note that the weighted state $\tau$ is a multilinear polynomial of its two inputs $\rho^{(0,1)}$.
Consider the algorithm in Fig.~\ref{fig:QSP}. Specifically, the pre-measurement output of the circuit is 
\begin{equation}
    \begin{aligned}
    \rho^\text{out} = \sum_{ii'jj'} &\sigma_{00} \rho^{(0)}_{ij} \rho^{(1)}_{i'j'} \ketbra{i0i'}{j0j'} + \sigma_{01}  \rho^{(0)}_{ij} \rho^{(1)}_{i'j'} \ketbra{i0i'}{j'1j} \\ + &\sigma_{10} \rho^{(0)}_{ij} \rho^{(1)}_{i'j'} \ketbra{i'1i}{j0j'}  + \sigma_{11} \rho^{(0)}_{ij} \rho^{(1)}_{i'j'} \ketbra{i'1i}{j'1j} \, .
    \end{aligned}
\end{equation}
Performing the measurement $M$ we generate the weighted state 
\begin{equation}
    \begin{aligned}
            \tilde{\tau} = &\sigma_{00} M_{00}\Tr(\rho^{(1)}) \rho^{(0)} + \sigma_{11}  M_{11} \Tr(\rho^{(0)}) \rho^{(1)} \\ + &\sigma_{01} M_{10} \rho^{(0)}\rho^{(1)} + \sigma_{10} M_{01} \rho^{(1)}\rho^{(0)} \, .
    \end{aligned}
\end{equation}
Hence by appropriately choosing the initial state $\sigma$ and measurement operator $M$ it is possible to produce states of the form Eq.~\eqref{eq:StatePoly} as claimed. More concretely, interpreting the coefficients $\alpha_{ij}$ as entries of a matrix $\alpha$ we require that
\begin{align}\label{eq:Alpha}
    \alpha = \sigma \odot M^T \odot \gamma^\text{in} \, .
\end{align} 
Here $\gamma^\text{in}$ is a matrix where the off-diagonal elements are input independent with $\gamma^{\text{in}}_{01}=\gamma^{\text{in}}_{10}=1$ and the diagonal entries depend on the input states with $\gamma^\text{in}_{00}=\Tr(\rho^{(1)})$ and $\gamma^\text{in}_{11}=\Tr(\rho^{(0)})$.

Multilinear polynomials of more than two input states, generalizing Eq.~\eqref{eq:StatePoly}, can be obtained by concatenating this algorithm. However, only a limited subset of all multilinear polynomials can be generated this way. In order to obtain any arbitrary multilinear polynomial, one can either randomize over instruments, see Appendix~\ref{ap:beyondnormal}, or forego concatenation and implement the transformation directly with an instrument that takes all inputs at once, see Appendix~\ref{ap:allinone}.

Ozols~\cite{ozols2015combine} describes an alternative algorithm for the task of QSP. In that approach there is no ancilla register. A unitary, that is a linear combination of identity and SWAP, is applied to the two states and one of the registers is traced out. The remaining register is in a proper quantum state that is of the form Eq.~\eqref{eq:StatePoly}. This algorithm also fits our general framework, but is less general in that it can only generate a small subset of states that our QSP algorithm can.

\medskip

A number of interesting families of weighted states may be generated with QSP when $\sigma$ is a physical density operator and $M$ is an Hermitian observable. Here we list several pertinent examples. 

\medskip

\paragraph*{Mixtures:} One does not need weighted states to prepare the mixture of two given states but here we show how that scheme fits in the larger framework of weighted states. Here we want $\tilde{\tau} = p \rho^{(0)} + (1-p) \rho^{(1)}$ for some probability $0<p<1$. 
This corresponds to $\alpha = \bigl( \begin{smallmatrix} p & 0\\ 0 & 1-p \end{smallmatrix}\bigr)$. 
We can achieve this by choosing $\sigma_A = \bigl( \begin{smallmatrix} p & 0\\ 0 & 1-p\end{smallmatrix}\bigr)$ and $M = \bigl( \begin{smallmatrix}1/\Tr(\rho^{(1)}) & 0\\ 0 & 1/\Tr(\rho^{(0)})\end{smallmatrix}\bigr)$. When the inputs are physical states with unit trace, the measurement $M$ is just the identity operator and so the environment register may simply be traced out.

\medskip

\paragraph*{Anti-commutator:} The weighted state $\tilde{\tau}=\{\rho^{(0)},\rho^{(1)}\}=\rho^{(0)}\rho^{(1)}+\rho^{(1)}\rho^{(0)}$ corresponds to $\alpha = \bigl( \begin{smallmatrix}0 & 1\\ 1 & 0\end{smallmatrix}\bigr)$. This may be prepared using $\sigma = \ketbra{+}{+}=(1/2)\bigl( \begin{smallmatrix}1 & 1\\ 1 & 1\end{smallmatrix}\bigr)$ and $M=2 X = 2 \bigl( \begin{smallmatrix}0 & 1\\ 1 & 0\end{smallmatrix}\bigr)$. We note that one-half of the anti-commutator with identical inputs $\rho$ is the weighted state $\tilde \tau= \rho^2$. In this special case, the algorithm reduces to established algorithms for computing Renyi entropies~\cite{Johri2017Entanglement,Yirka2021qubitefficient} and virtual state distillation~\cite{huggins2020virtual, czarnik2021qubit}.

\medskip

\paragraph*{Commutator:} The weighted state $\tilde{\tau}=[\rho^{(0)},\rho^{(1)}]=(\rho^{(0)}\rho^{(1)}-\rho^{(1)}\rho^{(0)})$ corresponds to $\alpha=\bigl( \begin{smallmatrix}0 & 1\\ -1 & 0\end{smallmatrix}\bigr)$. This may be prepared with $\sigma = \ketbra{+}{+}=(1/2)\bigl( \begin{smallmatrix}1 & 1\\ 1 & 1\end{smallmatrix}\bigr)$ and $M=2i Y =2\bigl( \begin{smallmatrix}0 & -1\\ 1 & 0\end{smallmatrix}\bigr)$.

\medskip

\paragraph*{Linear Combinations of Pure States:} A particularly promising application of the QSP algorithm is to prepare linear combinations of pure quantum states. In this case the weighted state is $\tau = | \psi \rangle \langle \psi |$ where $ | \psi \rangle = \alpha_0 |\psi^{(0)}\rangle + \alpha_1 | \psi^{(1)} \rangle$, corresponding to
\begin{equation}
    \alpha=\left( \begin{matrix} |\alpha_0|^2 & \frac{\alpha_0 \alpha_1^*}{\langle \psi_0|\psi_1 \rangle} \\ \frac{\alpha_1 \alpha_0^*}{\langle \psi_1|\psi_0\rangle} & |\alpha_1|^2 \end{matrix}\right) \, .
\end{equation}
Note that unlike previous algorithms we presented, in this case, as $\alpha$ depends on the overlap between the input states, the functional form of $g$ in Eq.~\eqref{eq:es_def_pure} depends on the input state.
In particular, one needs to know in advance, or estimate, the overlaps between the input states in order to specify the quantum instrument that prepares the desired weighted state.

We have freedom in how to pick the initial state $\sigma$ and measurement $M$. Suppose we take $\sigma$ to be the arbitrary pure state $\sigma = |\beta \rangle \langle \beta |$ with $| \beta \rangle = \beta_0 |0\rangle + \beta_1 | 1 \rangle$ then we need 
\begin{equation}\label{eq:Mlinearcombo}
    M =\left( \begin{matrix} \frac{|\alpha_0|^2}{|\beta_0|^2 \langle \psi_1 | \psi_1 \rangle} & \frac{\alpha_1 \alpha_0^*}{\beta_1 \beta_0^* \langle \psi_1 | \psi_0 \rangle } \\\frac{\alpha_0 \alpha_1^*}{\beta_0 \beta_1^* \langle \psi_0 | \psi_1 \rangle } & \frac{|\alpha_1|^2}{|\beta_1|^2\langle \psi_0 | \psi_0 \rangle} \end{matrix}\right) \, .
\end{equation}
Note that $M$ is a Hermitian operator. 
In Section~\ref{sec:sampling} we discuss how to choose $|\beta \rangle$ in order to minimize the sampling complexity.

We note that the denominator for the off diagonal elements in Eq.~\eqref{eq:Mlinearcombo} will vanish if $\ket{\psi_0}$ and $\ket{\psi_1}$ are orthogonal. Even if the denominator does not vanish, for close to orthogonal states it can become very small. As we will demonstrate in the error analysis in Section~\ref{sec:sampling} this leads to precision issues. First, the overlap $\langle \psi_0 | \psi_1 \rangle$ will have to be estimated with high precision.
Second, even if this overlap is known exactly the eigenvalues of $M$, i.e. the weights, will be large. This in turn increases the sampling complexity.
Hence, this method is not recommended for preparing linear combinations of orthogonal, or close to orthogonal, states. 

\medskip

One can prepare linear combinations of many states by iterating this method for preparing the linear combination of a pair of states.
This ability to take the linear combinations of states is expected to prove valuable for a number of applications since it can be used as a primitive to prepare arbitrary polynomials of quantum states. In particular, by taking the linear combination of powers of quantum states (generated via the QHP algorithm) one may approximately implement any function of the quantum state that may be expanded as a power series as
\begin{equation}
     \ket{\psi} \rightarrow  \ket{g(\psi)} = \sum_{i=1}^d g(\psi_i) \ket{i}
\end{equation}
where $g(\psi_i) = \sum_{k=1}^{K} \alpha_k (\psi_i)^k$. For example, this could be used to approximately implement the reciprocal operation $g(x) = \frac{1}{x}$ to amplify basis states with small amplitudes. Or, one could potentially use this method to implement the $g(x) = \tanh(x)$ activation function used to introduce nonlinearities into neural networks.

\zo{A polynomial of order $K$ of an $n$-qubit pure state can be implemented via a concatenation of QSP and QHP using $O(n K)$ 2-qubit gates.}
More generally the QSP, GQT, and QHP may be concatenated to implement nonlinear transformations of the form
\begin{equation}\label{eq:concat}
  \ket{\psi} \rightarrow  \ket{h(\psi)} = \sum_{k,l=1}^{KL} \sum_{i=1}^d\alpha_{kl} (\psi_i)^k (\psi_i^*)^l \ket{i}  \, .
\end{equation}
\zo{The number of two qubit gates required in this case scales as $O(n \chi^2)$ where $\chi = \text{max} \{ K, L \}$.}

\medskip
\zo{In Appendix~\ref{ap:AltLinearCombo} we discuss alternative methods for preparing linear combinations of states. The first of these is similar in spirit to the QSP method but creates the linear combination of multiple states in a single step (rather than via concatenation) with a single global measurement. This method is a more general than the QSP, capturing it as a special case. The second method we detail is of different spirit to the other algorithms detailed here (i.e. it does not fall naturally within the weighted state framework) and uses a combination of Hadamard tests and classical post-processing. This method requires shorter depth circuits and thus is more appropriate for near-term hardware as shown in Appendix~\ref{ap:HardwareImp}.}

\medskip

Physically measurements are associated with normal operators. If we require $M$ to be a normal operator in the QSP circuit, then it follows from Eq.~\eqref{eq:Alpha} that not all $\alpha$ matrices can be obtained. In Appendix~\ref{ap:RealizableStates} we provide a detailed classification of the classes of weighted states that can be realised when $M$ is a normal operator. 
To go beyond, as previously noted, we can express arbitrary operators as sums of normal operators $N_l$ via $M = \sum_l c_l N_l$. The right hand side of Eq.~\eqref{eq:map} now has a summation, which means that we need to operate multiple quantum instruments to obtain the weighted state. As an example of weighted states that can only be prepared via QSP through such means consider products of density matrices, i.e. $\tau=\rho^{(0)}\rho^{(1)}$.

\section{Sampling complexity analysis}\label{sec:sampling}

We are interested in the expectation value of an Hermitian operator $O$ in a weighted state $\tau$, i.e. $\Tr[\tau O]$. 
Due to Eq.~\eqref{eq:map} this quantity is nothing but the expectation value of $O\otimes M\otimes I$ in the state $\rho^\text{out}$ and can be estimated by repeatedly running the quantum instrument of Fig.~\ref{fig:es} and then measuring the operator $O$ on the quantum output, i.e. the system register. This estimator takes the form
\begin{align}
    \widehat{\mathcal{O}} = \frac{1}{s} \sum_i^{s} \lambda_{m(i)} \mu_{o(i)}
\end{align}
where $\lambda_{m(i)}$ and $\mu_{o(i)}$ are the random outcomes of $M$ and $O$ measurements at the $i$'th run of the circuit, respectively, and $s$ is the total number of runs. Note that $\widehat{\mathcal{O}}$ is an unbiased estimator since it estimates a quantum expectation value in the standard way: 
\begin{align}
    E[\widehat{\mathcal{O}}] = \Tr[\rho^\text{out} (O \otimes M\otimes \mathbb{I})  ] = \Tr[\tau O] 
\end{align}
where $E[\cdot]$ denotes the expectation value.

The variance of $\widehat{\mathcal{O}}$ is
\begin{equation}
    \label{eq:Var}
    \text{Var}(\widehat{\mathcal{O}}) = \frac{1}{s} \left( \Tr[\rho^{\mbox{\tiny out}} (O \otimes M\otimes \mathbb{I})^2 ] - \Tr[\tau O]^2\right) \, .
\end{equation}
Note that, in general, the variance of an observable can not be expressed in terms of the weighted state $\tau$ alone. Consequently, two instruments that implement the same transformation and hence prepare the same weighted state, can result in estimators with different variances. 
Since variance is inversely proportional to sampling complexity we would like to find instruments that minimize it. As is clear from Eq.~\eqref{eq:Var}, the variance manifestly depends on the measurement $O$ performed on the weighted state and so to compare the sampling complexity of different instruments it is desirable to derive an operator independent bound on the variance. 
Assuming that $|| O||_\infty \leq 1$, it follows that 
\begin{equation}\label{eq:ErrorBoundVar}
  \text{Var}(\widehat{\mathcal{O}})  \leq \frac{\Tr[\rho^\text{out} ( \mathbb{I} \otimes M\otimes \mathbb{I})^2 ] }{s} \, .
\end{equation}
We use Eq.~\eqref{eq:Var} (Eq.~\eqref{eq:ErrorBoundVar}) to compute (bound) the number of samples needed for QHP, GQT, and QSP algorithms in order to achieve a desired precision.
Here we restrict our scope to weighted states that are prepared using quantum instruments whose inputs are physical states\footnote{Note that if the input states are weighted states themselves, one needs to take into account the quantum instruments associated with them in order to analyze the variance. 
This can be achieved by treating the concatenated quantum instruments as a single quantum instrument with more registers.} and for which $M$ is a Hermitian operator. We also suppose that the system observable $O$ is Hermitian. Here we summarise our findings. The more detailed and general analysis can be found in Appendix.~\ref{ap:Errors}.

\subsection{Quantum Hadamard Product}

For the Quantum Hadamard Product algorithm the variance takes the form
\begin{align}\label{eq:QHPerror}
    \text{Var}[\hat{\mathcal{O}}] 
    &=\frac{1}{s}(\Tr[\tau O^2] - \Tr[\tau O]^2)
\end{align}
Thus the variance has the form of a standard quantum observable, except that in this case $\tau$ may be a sub-normalized state with $\Tr[\tau]\le1$. This expression, Eq.~\eqref{eq:QHPerror}, also holds when the QHP circuit is applied iteratively to generate higher order powers/polynomials of the input states. 

\begin{figure}[t]
    \centering
    \includegraphics[width=\columnwidth]{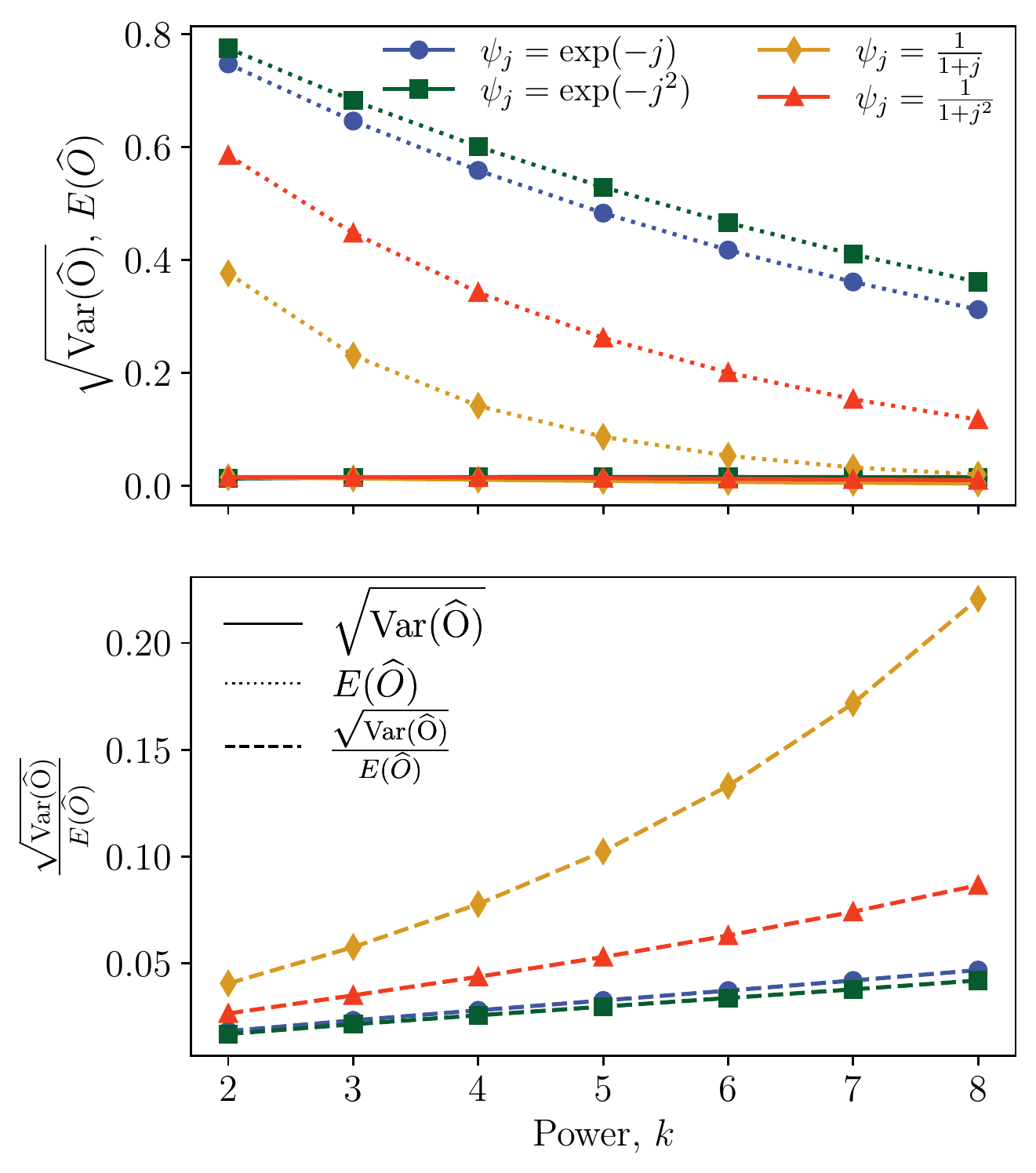}
    \caption{\textbf{Error analysis for powers of states.} Here we plot the absolute sampling error $\sqrt{\text{Var}(\widehat{\mathcal{O}})}$ (solid), the analytically computed expectation value $\text{E}(\widehat{\mathcal{O}})$ (dotted) and the relative error $\sqrt{\text{Var}(\widehat{\mathcal{O}})}/\text{E}(\widehat{\mathcal{O}})$ (dashed) after running the powers of state circuit for the normalized states $|\psi\rangle \propto \sum_j \psi_j |j\rangle$, with the functions $\psi_j$ indicated in the legend, as a function of power $k$. In all cases we consider an $n=6$ qubit state $\ket{\psi}$, we measure the all zero projector $O = | 0 \rangle \langle 0 |$ and suppose $s=1000$ shots are used.}
\label{fig:PowerError}
\end{figure}

In Fig.~\ref{fig:PowerError} we plot the absolute and relative sampling errors when applying the QHP algorithm for generating increasing powers of different input states. While the additive error (solid lines) remains approximately constant as $k$ is increased, the relative error 
\begin{equation}
    \frac{ \text{Var}(\sqrt{\widehat{\mathcal{O}})}}{ \Tr[\tau O]} = \sqrt{ \frac{1}{s} \left(  \frac{\Tr[\tau O^2]}{ \Tr[\tau O]^2} - 1 \right)}
\end{equation}
typically increases dramatically with power (dashed lines). This follows from the fact that the amplitudes of $|\psi \rangle$ are typically less than $1$ (for proper quantum states) and so $\Tr[\tau]=\sum_i |\psi_i|^{2k}$ typically decreases exponentially with $k$ (dotted lines).

\subsection{Generalized Quantum Transpose}

For the Generalized Quantum Transpose circuit shown in Fig.~\ref{fig:qhp}, the variance is given by 
\begin{equation}
\label{eq:varGQT}
    \text{Var}(\widehat{\mathcal{O}})= \frac{1}{s} \left( \Tr\left[D(\sigma) O^2\right] - \Tr\left[ \left( \sigma \odot \rho^{T} \right) O \right]^2 \right)
\end{equation}
where $D(\sigma)$ is the dephased version of the ancilla state $\sigma$, i.e. $D(\sigma) = \sum_i \sigma_{ii} |i\rangle\langle i|$, and we used the fact that $\text{SWAP}^2=\mathbb{I}$.

We recall that the Generalized Quantum Transpose may be used to compute expectation values of the transpose of the state $\rho$ if $\sigma$ is chosen to be the plus state $| + \rangle \langle + |$ and the observable $d O$ is measured. The factor of $d$ compensates for the normalization factor in $\rho_{|+\rangle\langle+|}^{(T)} = \frac{1}{d} \rho^{T}$ (where here we use the notation for the generalized transpose introduced in Eq.~\eqref{eq:QHP}). While this is one possible use of GQT, in this case the variance is given by 
\begin{equation}
    \text{Var}(\widehat{\mathcal{O}})= \frac{1}{s} \left( d \Tr\left[O^2\right] - \Tr\left[ \rho^{T} O \right]^2 \right) \, ,
\end{equation}
such that the sampling error scales with $d$ and so exponentially in the number of qubits of the input system. Thus for large scale problems GQT is not efficient. However, if GQT is used to implement the partial transpose to a $d_s$ dimensional subsystem then the variance will scale as $d_s$. Hence even for large scale problems it should be possible to implement the partial transpose of a constant sized subsystem.  

\begin{figure}[t!]
    \centering
    \includegraphics[width=\columnwidth]{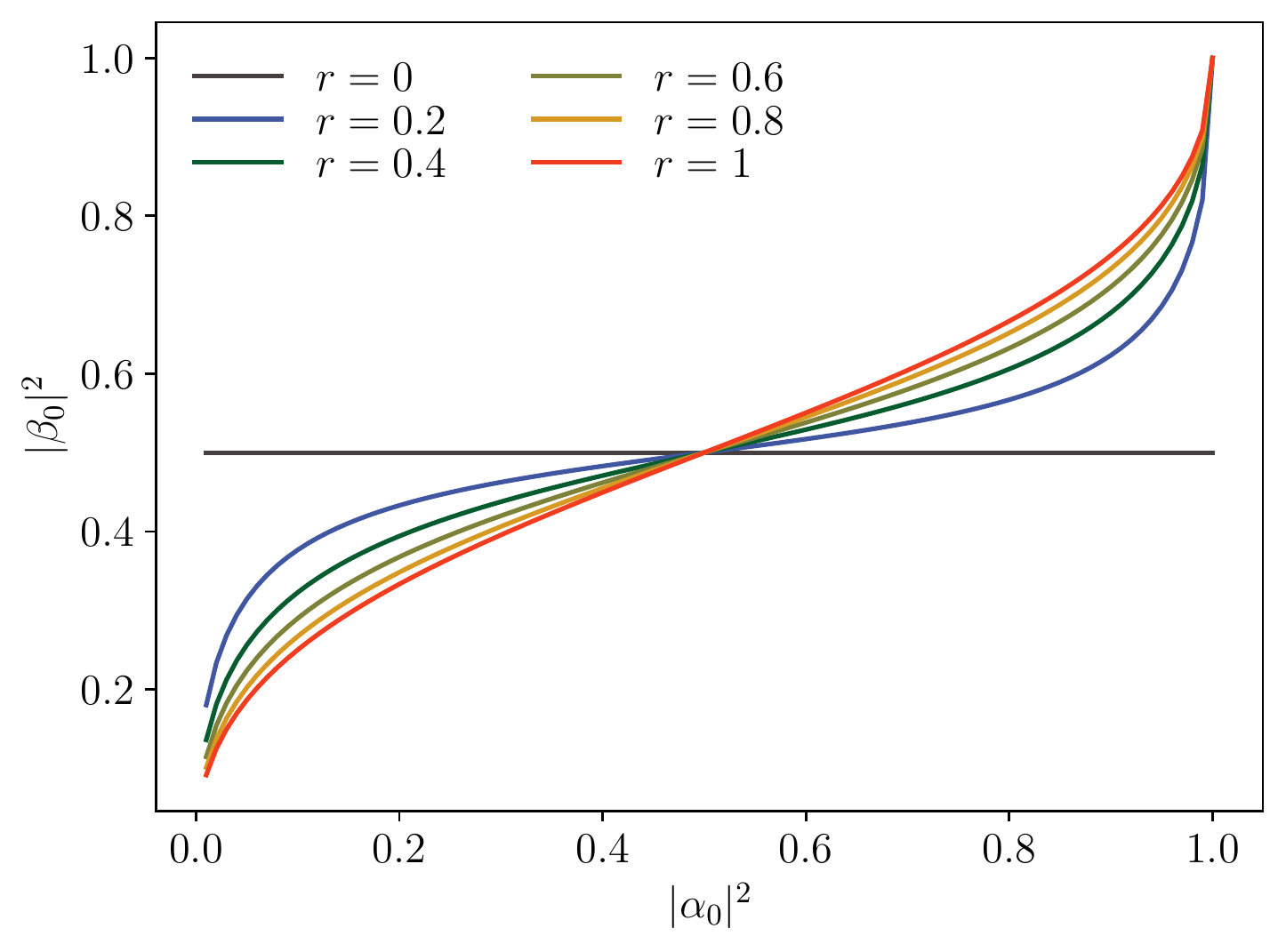}
    \caption{\textbf{Optimum choice in $\beta$ parameter.} We plot $\beta_0^{\rm opt}$, the optimum $\beta_0$ parameter that minimises $\mathcal{B}_{\text{Var}}$, the upper bound on the error for the linear combinations of states circuit, as a function of $|\alpha_0|^2$ for different overlaps $r$.}
    \label{fig:optstrat}
\end{figure}

\begin{figure*}[t!]
    \centering
    \includegraphics[width=0.98\textwidth]{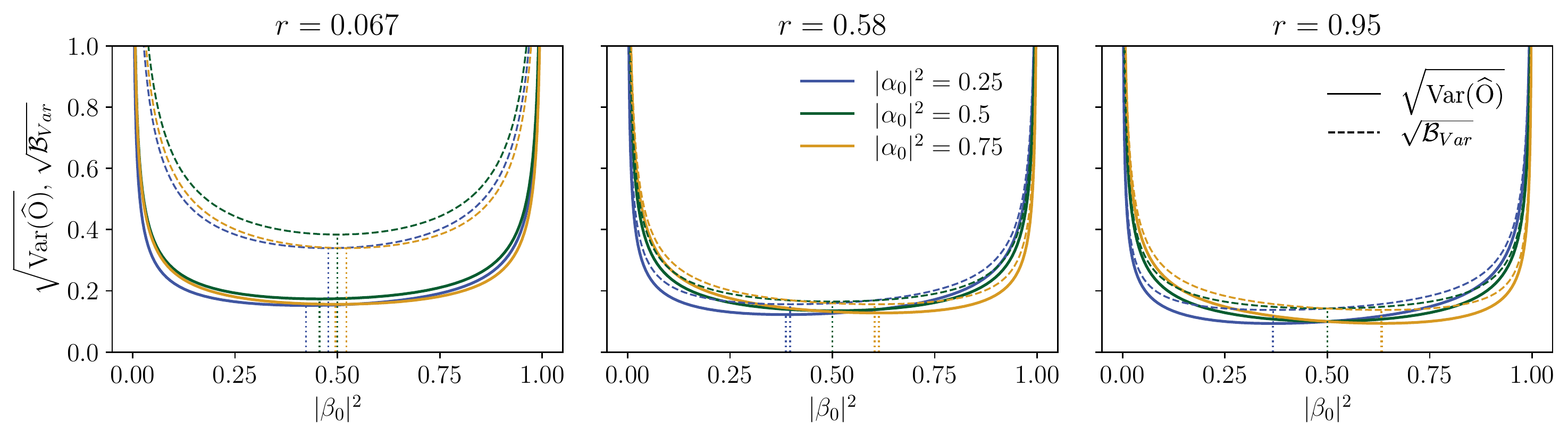}
    \caption{\textbf{Error Analysis for Linear Combination of Unitaries.} Here we plot the the actual standard deviation of the estimator, $\sqrt{\text{Var}(\widehat{\mathcal{O}})}$, (solid) and its upper bound $\sqrt{\mathcal{B}_{\text{Var}}}$ (dashed) for the linear combination of states algorithm for an $n=6$ qubit circuit evaluated using $N = 100$ shots when $O$ is a random separable measurement operator. We plot the results for pairs of randomly generated states with a small overlap ($r = 0.067$, left), median overlap ($r = 0.58$, middle) and large overlap ($r = 0.95$, right) for three different choices in amplitude: $\alpha_0 = 0.25$ (blue),  $\alpha_0 = 0.5$ (green) and $\alpha_0 = 0.95$ (yellow). The dotted lines indicate the value of $|\beta_0|^2$ corresponding to the minimum error and minimum of the bound respectively, with the close agreement between these two minima indicating that the optimum $|\beta_0|^2$ determined from the bound, Eq.~\eqref{eq:optBeta}, is close to the true optimum. In agreement with Fig.~\ref{fig:optstrat} the optimal $|\beta_0|^2$ value is close to $0.5$ for small $r$ but has a stronger dependence on $\alpha_0$ for larger $r$ values. }\label{fig:LinError}
\end{figure*}

\subsection{Linear combination of states}\label{sec:LinearComb}

Here we focus on analysing the sampling complexity of the linear combination of states algorithm - see Appendix~\ref{ap:Errors} for a presentation of the errors in the general case of the quantum state polynomial circuit. Using Eq.~\eqref{eq:ErrorBoundVar}, one can show that for the linear combination of states algorithm the variance may be bounded as 
\begin{equation}
\begin{aligned}\label{eq:boundLinCombo}
       \text{Var}(\widehat{\mathcal{O}})  \leq \frac{1}{s} \bigg(  &|\alpha_0|^2 \left(\frac{|\alpha_0|^2}{|\beta_0|^2}  + \frac{|\alpha_1|^2}{q_1r} \right)  \\ + &|\alpha_1|^2 \left( \frac{|\alpha_0|^2}{|\beta_0|^2 r} + \frac{|\alpha_1|^2}{|\beta_1|^2} \right) \bigg) := \mathcal{B}_{\text{Var}} \, ,
\end{aligned}
\end{equation}
where $r = | \langle \psi_0 | \psi_1 \rangle|^2$ is the overlap between the input states. 
As remarked earlier, for this algorithm there is a freedom in how the ancilla state and measurement operator may be chosen. We can use the above bound to approximately determine the optimum $\beta_0$ value which minimises the variance of the estimator. Specifically, we find that the $\beta_0$ value that minimizes $\mathcal{B}_{\text{Var}}$ is 
\begin{equation}\label{eq:optBetaVal}
    \beta_0^{\rm opt} (p, r) = \sqrt{\frac{p - p^2 + p^2 r + r^{-1} h(p,r)}{2 p - 2 p^2 + r - 2 p r + 2 p^2 r + 
   2 r^{-1}h(p,r)} }
\end{equation}
where we use the shorthand $p := |\alpha_0|^2$ and define
\begin{equation}
\small 
     h(p,r) := \sqrt{(p^2 - p) (-p + p^2 - r + 2 p r - 2 p^2 r - p r^2 + p^2 r^2)}  \, . 
\end{equation}
We plot $\beta_0^{\rm opt}$ against $|\alpha_0|^2$ in Fig.~\ref{fig:optstrat}. For states with small overlaps, $r \rightarrow 0$, the approximately optimum strategy is to use $\beta_0 = 1/\sqrt{2}$ for any choice in $\alpha_0$, whereas, for states with large overlaps, $r \rightarrow 1$, the optimum $\beta_0$ value increases monotonically with increasing $\alpha_0$. 

In Fig.~\ref{fig:LinError} we plot the exact variance of the estimator (for the analytic expression see Appendix~\ref{ap:Errors}) and the bound on the variance $\mathcal{B}_{\text{Var}}$ from Eq.\eqref{eq:boundLinCombo} as a function of $\beta_0$, for different overlaps $r$ and weights $\alpha_0$. While the bound is not tight it is useful in coming up with strategies to minimize the variance of the estimator. In particular, the optimum $\beta_0$ given by Eq.~\eqref{eq:optBetaVal}, found by minimizing the bound Eq.\eqref{eq:boundLinCombo}, closely agrees with the true optimum (found by numerically minimising the true variance for the given examples). As expected given that the algorithm breaks down for orthogonal input states, both the bound and exact variance diverge for input states with vanishing overlap ($r \rightarrow 0$).

\section{Discussion}\label{sec:discussion}

In this work we introduced a framework for implementing nonlinear transformations in quantum computers by associating so-called weighted states with the output of quantum instruments. More specifically, weighted states are  quantum objects describing the output of an operational procedure involving quantum circuits, measurements and classical post-processing. While playing a similar role to standard density matrices, weighted states are liberated from the constraints of positivity, Hermiticity and normalization and hence can be generic functions of input states.

We have introduced three algorithms for implementing nonlinear functionals of the elements of a set of quantum states. The Quantum Hadamard Product algorithm takes two states $\rho^{(0)}$ and $\rho^{(1)}$ as inputs and generates a state $\rho^{(0)}\odot \rho^{(1)}$ as an output. That is, QHP outputs a weighted state where the elements of $\rho^{(0)}$ and $\rho^{(1)}$ have been multiplied in the computational basis. The Generalised Quantum Transpose implements the transpose of an operator $\rho$ in the computational basis, with the elements of $\rho^T$ reweighted by the elements of an operator $\sigma$. The Quantum State Polynomial algorithm takes $\rho^{(0)}$ and $\rho^{(1)}$ as inputs and outputs the polynomial $\alpha_{00} \rho^{(0)}+ \alpha_{11}  \rho^{(1)} + \alpha_{01} \rho^{(0)} \rho^{(1)}+ \alpha_{10}  \rho^{(1)} \rho^{(0)}$. When applied to pure states, iterative applications of QHP, GQT, QSP can be used to generate arbitrary polynomials of the amplitudes of a set of pure states. 

In Appendix~\ref{ap:HardwareImp} we show results from a proof of principle implementation of QSP and QHP on IBMQ-Bogota. While the implementations correctly capture the qualitative effect of performing nonlinear transformations of quantum states, the low CNOT and qubit reset fidelities lead to non-negligible deviations from the expected results. 
However, it is important to note that, due to their amenability to qubit resets~\cite{Yirka2021qubitefficient, czarnik2021qubit}, these algorithms do not require large numbers of qubits. In particular, QSP, GQT, and QHP can be concatenated and iterated an arbitrarily many times with at most three times the number of qubits of the input states.
Thus a polynomial of order $p$ of an $n$-qubit pure state can be implemented using at most $3n$ qubits (for any $p$).

It is important to note that distinct quantum instruments can implement the same transformation with differing complexities. 
In other words, in this framework, the complexity associated with a transformation is not an inherent property of the transformation. However, one can ask what is the optimal quantum instrument for a given transformation. 
A promising setting in which this question might be tractable is when the cost of implementing unitaries is neglected, in other words we are only concerned with minimizing the sampling complexity. 
How exactly to do this optimization is an open question;
however, in Section~\ref{sec:LinearComb} we were able to do a partial optimization for one of our algorithms (namely, QSP as applied to implementing linear combinations of states). 

In this paper we focused on primitives that implement intuitive transformations and require relatively simple quantum circuits for their implementation. These primitives can be concatenated to yield complex nonlinear transformations of input states. 
While concatenation is qubit-efficient, in Appendix~\ref{ap:concatenationcost} we provide examples of cases where a transformation can be implemented more efficiently directly. Thus it would be valuable to have a method for directly designing quantum instruments for a target transformation.
More generally, it would be interesting to investigate the breadth of applicability of the weighted state methodology by quantifying the full set of transformations that can be implemented within this framework.

\section{Acknowledgements}
We thank Denis Aslangil, Lukasz Cincio, Patrick Coles, Yen Ting Lin, Robert Lowrie, Rolando Somma, and Burak \c{S}ahino\u{g}lu for useful discussions. We further thank Mark Wilde for bringing the quantum teleportation link to our attention.
This work was supported by the Beyond Moore's Law project of the Advanced Simulation and Computing Program at Los Alamos National Laboratory (LANL) managed by Triad National Security, LLC, for the National Nuclear Security Administration of the U.S. Department of Energy under contract 89233218CNA000001.
Research presented in this article was also supported by the Laboratory Directed Research and Development program of LANL. 
ZH acknowledges support from the Mark Kac Fellowship. 
Part of this work was completed while NC was a participant in the 2021 Quantum Computing Summer School at LANL, sponsored by the LANL Information Science \& Technology Institute.

\bibliography{main}

\onecolumngrid

\clearpage

\appendix

\vspace{0.5in}

\begin{center}
	{\Large \bf Appendices} 
\end{center}

\section{Quantum instruments}

In this section we elaborate on the quantum instruments associated with the quantum Hadamard product, quantum generalized transpose, and density matrix polynomial algorithms and provide alternative derivations of these algorithms from this perspective.

\subsection{Quantum Hadamard Product}

The quantum instrument for the QHP is shown in Fig.~\ref{fig:qhp}. Let $x$ be the outcome of a measurement in the computational basis. Given this classical output of the quantum instrument, the associated quantum output is proportional to:
\begin{align}
    \mathcal{E}_x(\rho^{(0)}\otimes\rho^{(1)}) = \sum_{i,j} \rho^{(0)}_{ij} \rho^{(1)}_{i\oplus x,j\oplus x} \ket{i}\!\bra{j}
\end{align}
The weighting in this case corresponds to post-selection; only $x=0$ has unit weight and all other outcomes have zero weight. 
\begin{align}
    \tilde{\tau} &= \sum_x \delta_{x,0} \mathcal{E}_x(\rho^{(0)}\otimes\rho^{(1)}) \\
    &= \sum_{i,j} \rho^{(0)}_{ij} \rho^{(1)}_{i,j} \ket{i}\!\bra{j} = \rho^{(0)} \odot \rho^{(1)}
\end{align}
Note that one can generate a large family of weighted states by projecting onto states other than $\ket{0}$ and/or nontrivially weighting all outcomes of the measurement.  

\subsection{Generalized Quantum Transpose}

The quantum instrument for the GQT is shown in Fig.~\ref{fig:gqt}. 
In practice the measurement of SWAP operator is implemented by measuring pairs of qubits in the Bell basis as seen in Fig~\ref{fig:SWAP}~\cite{cincio2018learning}. This can be achieved by first acting with a CNOT on qubit pairs and then Hadamard gates on the control qubits, and finally measuring all qubits in the computational basis. Let $x$ and $y$ be the binary strings that are the outcome of the measurements in the two registers where SWAP operator is measured. Given this classical output of the quantum instrument, the associated quantum output is proportional to:
\begin{align}
    \mathcal{E}_{x,y}(\sigma \otimes \rho) =  \frac{1}{2^{n}} \sum_{ij} \sigma_{i,j} \rho_{i\oplus y,j\oplus y} (-1)^{x\cdot(i\oplus j)} \ketbra{i}{j}
\end{align}
where $n$ is the number of qubits in each input register.
The weighted state is obtained by weighting the quantum outcomes of the instruments as:
\begin{align}
    \tilde{\tau} &= \sum_{x,y} (-1)^{x\cdot y} \mathcal{E}_{x,y}(\sigma\otimes \rho) \\
    &=  \sum_y \sum_{ij} \sigma_{i,j} \rho_{i\oplus y,j\oplus y} \left(\frac{1}{2^{n}} \sum_x (-1)^{x\cdot(i\oplus j \oplus y)}\right) \ketbra{i}{j} \\
    &= \sum_{ij}\sigma_{i,j} \rho_{j,i} = \sigma \odot \rho^T
\end{align}
where going from the second to the last line we used the fact $ \sum_x (-1)^{x\cdot(i\oplus j\oplus y)} = 2^{n} \delta_{y,i\oplus j}$ and $i\oplus i \oplus j = j$. All summations are modulo 2.

\subsection{Quantum State Polynomial}

The quantum instrument for QSP is shown in Fig.~\ref{fig:QSP}. Let $\ket{\psi_{l}}$ be the state that correspond to the $l$'th outcome of the $M$ measurement. Given this classical output of the quantum instrument, the associated quantum output is proportional to:
\begin{align}
    \mathcal{E}_{l}(\rho^{(0)}\otimes \rho^{(1)}) = \sum_{i,j}  \sigma_{ij}\braket{j|\psi_l}\braket{\psi_l|i} \left[\delta_{ij}\Tr(\rho^{(i\oplus 1)}) \rho^{(i)} + (1-\delta_{ij}) \rho^{(i)} \rho^{(j)} \right]
\end{align}
The weighted state is obtained by weighting the quantum outcomes of the instrument as:
\begin{align}
    \tilde{\tau} &= \sum_l \lambda_l \mathcal{E}_{l}(\rho^{(0)}\otimes \rho^{(1)})\\
    &=\sum_{i,j} \sigma_{ij} \bra{j}\left(\sum_l \lambda_l \ket{\psi_l}\! \bra{\psi_l} \right)\ket{i}\left[\delta_{ij}\Tr(\rho^{(i\oplus 1)}) \rho^{(i)} + (1-\delta_{ij}) \rho^{(i)} \rho^{(j)} \right] \\
    &= \sum_{ij} (\sigma \odot M^T)_{ij}\left[\delta_{ij}\Tr(\rho^{(i\oplus 1)}) \rho^{(i)} + (1-\delta_{ij}) \rho^{(i)} \rho^{(j)} \right] \\
    &= \sigma_{00} M_{00}\Tr(\rho^{(1)}) \rho^{(0)} + \sigma_{11}  M_{11} \Tr(\rho^{(0)}) \rho^{(1)} +\sigma_{01} M_{10} \rho^{(0)}\rho^{(1)} + \sigma_{10} M_{01} \rho^{(1)}\rho^{(0)} 
\end{align}
where going from the second to the last line we used the fact $M = \sum_l \lambda_l \ket{\psi_l}\!\bra{\psi_l}$.

\subsection{Going beyond normal measurement operators}
\label{ap:beyondnormal}

Let us consider a set of quantum instruments $\mathcal{I}_k(\sigma, U, N_k)$, labeled by $k$, that share the same ancilla state $\sigma$ and that apply the same unitary $U$, but differ in terms of the measurement operators $M$ now referred to as $N_k$ to emphasize they are normal. 
Data from such quantum instruments can be compounded to emulate a fictitious quantum instrument $\mathcal{I}_k(\sigma, U, \bar M)$ with a non-normal measurement operator $\overline{M}=\sum_k c_k N_k$. 
This follows from Eq.~\eqref{eq:map} via linearity. That is, as 
\begin{align}
\label{eq:Msum}
     \sum_k c_k \Tr[\rho^\text{out} (O \otimes N_k \otimes \mathbb{I})] &=  \Tr[\rho^\text{out} (O \otimes \sum_k c_k N_k \otimes \mathbb{I})] =\Tr[\rho^\text{out} (O \otimes \overline{M} \otimes \mathbb{I})] 
\end{align}
it follows that $\sum_k c_k \mathcal{I}_k(\sigma, U, N_k) \cong \mathcal{I}_k(\sigma, U, \bar M) $.
It is easy to see that this allows us to effectively implement any $M$ since any operator can be written as the sum of its Hermitian and anti-Hermitian components via
\begin{align}
    \overline{M} &= \frac{1}{2}(\overline{M}+\overline{M}^\dagger)+\frac{1}{2}(\overline{M}-\overline{M}^\dagger) \, .
\end{align}
As both parts are normal, we could pick two instruments $N_{0}=(\overline{M} + \overline{M}^\dagger)$ and $N_{1}=(\overline{M} -  \overline{M}^\dagger)$ with $c_0=c_1=1/2$ to achieve our goal. 
We emphasize that this example is only intended as a proof of existence and not a recipe. 
There are an infinite number of ways to satisfy Eq.~\eqref{eq:Msum}, and some ways may result in better sampling complexity than others.

\medskip

However, such non-normal measurement operators are not strictly necessary for full generality. This is because given an instrument $\mathcal{I}(\sigma, U, \overline{M})$, where $\overline{M}$ is a non-normal operator it is possible to define an alternative instrument $\mathcal{I}(\sigma \otimes \sigma', U \otimes \mathbb{I}, M)$, where $M$ is a normal operator, and the alternative instrument implements the same transformation as the original. 
 
\begin{figure}
    \centering
    \includegraphics[scale=1.0]{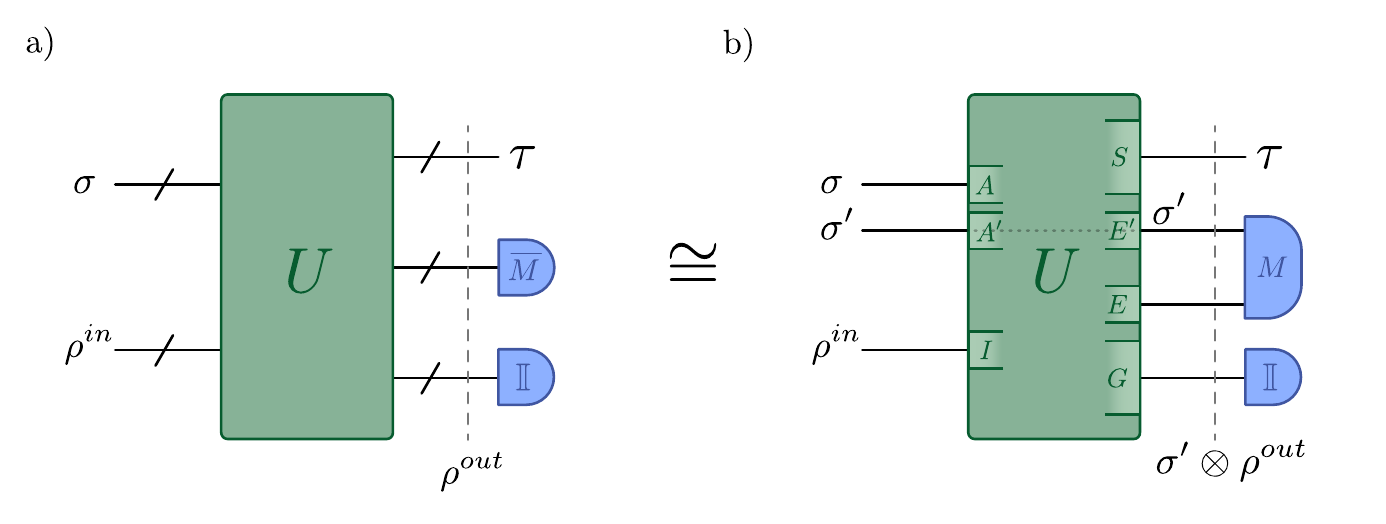} \caption{\textbf{Capturing a non-normal measurement within the standard weighted state framework.} Here we show two equivalent quantum instruments $\mathcal{I}(\sigma_A, U_{AI}, \overline{M}_{E})$ and $\mathcal{I}(\sigma_{A} \otimes \sigma'_{A'}, U_{AI} \otimes \mathbb{I}_{A'}, M_{E'E})$. }
    \label{fig:arbitrary_M}
\end{figure}

To see how let us, without loss of generality, assume $c_k>0$ and $\sum_k c_k=1$. The alternative quantum instrument $\mathcal{I}(\sigma_{A} \otimes \sigma'_{A'}, U_{AI} \otimes \mathbb{I}_{A'}, M_{E'E})$, where we now include subscripts to explicitly denote the relevant subsystem, shown in Fig.~\ref{fig:arbitrary_M}. There is an additional ancilla register initially prepared in the state $\sigma'_{A'} =\sum_k c_k \ketbra{k}{k}$. The unitary is the same as before and does not act on this new ancilla register. The environment register on the output now includes the added ancilla register. The measurement operator is given by ${M}_{E'E}=\sum_k \ketbra{k}{k}_{E'}\otimes {(N_k)}_E$, which is normal by construction. The quantum state before the measurement is ${\rho'}^\text{out}_{SEE'G} = \rho^\text{out}_{SEG} \otimes \sigma'_{E'}$ and so we have 
\begin{align}
\label{eq:arbitrary_M}
  \Tr_{SEE'G}[ (\rho^\text{out}_{SEG} \otimes \sigma'_{E'} ) (O_S \otimes  M_{EE'} \otimes \mathbb{I}_G)] &=  \Tr_{SEG}[\rho_{SEG}^\text{out} (O_S \otimes \sum_k c_k {(N_k)}_E \otimes \mathbb{I}_G)] \\ &= \Tr_{SEG}[\rho_{SEG}^\text{out} (O_S \otimes \overline M_E \otimes \mathbb{I}_G)] \, .
\end{align}
Thus $\mathcal{I}(\sigma_A, U_{AI}, \overline{M}_{E}) \cong \mathcal{I}(\sigma_{A} \otimes \sigma'_{A'}, U_{AI} \otimes \mathbb{I}_{A'}, M_{E'E})$ as claimed. Note that this construction is equivalent to randomly sampling quantum instruments with measurement operator $N_k$ with probability $c_k$, but the formulation in terms of a single instrument allows us to simplify notation for the rest of the discussion. 

\section{Alternative methods for preparing a linear combination of states}\label{ap:AltLinearCombo}

Given $L+1$ states $\{\ket{\psi_i}\}_{i=0}^{L}$, our goal is to simulate the effect of having access to a superposition of these states:
\begin{align}
    \label{eq:LCS}
    \ket{\Phi} &= \sum_{l=0}^{L} \alpha_l \ket{\phi_l} \, . 
\end{align}
The states $\{\ket{\phi_i}\}_{i=0}^{L}$ do not form an orthonormal basis, and $\ket{\Phi}$ does not have to be normalized. In the main text we describe a method for generating such linear combinations of states via repeated applications of the Quantum State Polynomial algorithm. Here we describe three alternative methods for this task. 

\subsection{`All-At-Once' weighted state method}
\label{ap:allinone}
In this section we describe a weighted state methodology where the linear combination of states is generated in a single step via a global measurement. This approach is more general than the iterative application of QSP algorithm, capturing it as a special case. 

\medskip

\begin{figure}
\centering
\includegraphics[]{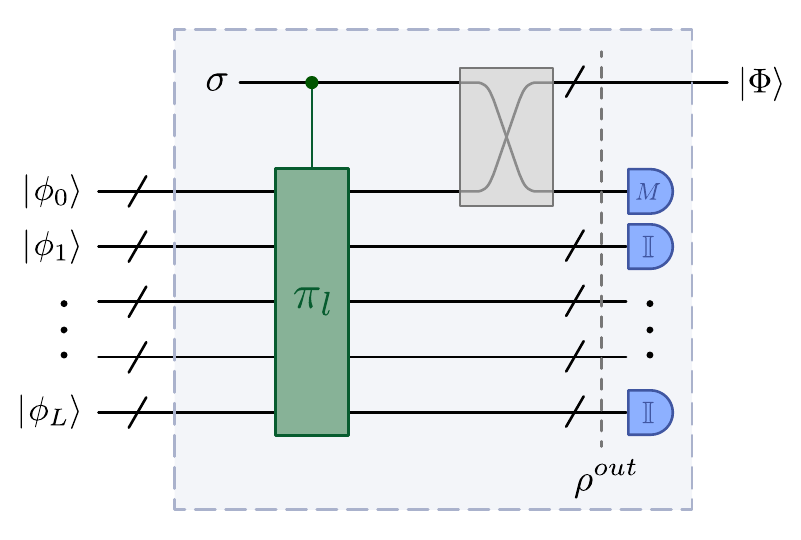}
\caption{\textbf{Direct Linear combination of states algorithm.} Here we show a circuit to directly compute the linear combination of many states. $\hat{\pi}_l$ is a permutation that maps the $l$'th state to the $0$'th state. The final swap does not need to be implemented but rather is included to match the register labelling convention chosen in Fig.~\ref{fig:es}.}
\label{fig:lcs}
\end{figure}

Consider the quantum circuit shown in Fig.~\ref{fig:lcs}. Here $\sigma = |\beta \rangle \langle \beta |$ where
\begin{align}
   |\beta\rangle  &= \sum_{l=0}^L \beta_l \ket{l}  \, .
\end{align}
The controlled unitary is given by
\begin{align}
    \ket{l}\!\bra{l}\otimes \hat{\pi}_l
\end{align}
where $\hat{\pi}_l$ is a permutation that maps the $l$'th state to the $0$'th state
\begin{align}
    \hat{\pi}_l \ket{\phi_0}\otimes\ket{\phi_1}\otimes\dots\otimes\ket{\phi_L} = \ket{\phi_l}\otimes\ket{\phi_{\pi_l(1)}}\dots \otimes \ket{\phi_{\pi_l(L)}} 
\end{align}
i.e. $\pi_l(0)=l$ for the state labels\footnote{One could equivalently see this operation as mapping the $0$'th register to the $l$'th register.}. Beyond this specification, we leave the permutations $\pi_l$ underspecified. 
The state at the end of this operation is given by
\begin{align}
    \sum_{l=0}^L \beta_l \ket{l} \otimes \ket{\phi_l} \bigotimes_{k=1}^L \ket{\phi_{\pi_l(k)}}
\end{align}
The next step is to trace out all but the top two registers. To do so, we first write down the density operator for the pure state of all the registers given above
\begin{align}
    \sum_{l,l'=0}^L \beta_l \beta_{l'}^* \ket{l}\!\bra{l'} \otimes \ket{\phi_l}\!\bra{\phi_{l'}} \bigotimes_{k=1}^L \ket{\phi_{\pi_l(k)}}\!\bra{\phi_{\pi_{l'}(k)}} \, .
\end{align}
Then, tracing out the registers 1 through $L$ we get
\begin{align}
    \sum_{l,l'=0}^L \beta_l \beta_{l'}^* \prod_{k=1}^L \braket{\phi_{\pi_{l'}(k)}|\phi_{\pi_l(k)}} \ket{l}\!\bra{l'} \otimes \ket{\phi_l}\!\bra{\phi_{l'}} \, .
\end{align}
Next we swap the top two registers and compute the expectation value of an operator $M$ on the lower register and output the top register. This leaves us with a weighted state given by
\begin{align}
    \sum_{l,l'=0}^L \beta_l \beta_{l'}^* \braket{l'|M|l} \prod_{k=1}^L \braket{\phi_{\pi_{l'}(k)}|\phi_{\pi_l(k)}}  \otimes \ket{\phi_l}\!\bra{\phi_{l'}}
\end{align}
We want this to match the density operator 
\begin{align}
    \ket{ \Phi}\!\Bra{ \Phi} &= \sum_{l,l'=0}^L \alpha_l \alpha_{l'}^* \ket{\phi_l}\!\bra{\phi_{l'}}
\end{align}
Matching term by term we demand
\begin{align}
    \label{eq:A}
    \braket{l'|M|l} = \frac{\alpha_l \alpha_{l'}^*}{\beta_l \beta_{l'}^*\prod_{k=1}^L \braket{\phi_{\pi_{l'}(k)}|\phi_{\pi_l(k)}}}
\end{align}
We remark that $M$ is Hermitian, i.e. $M_{ll'}^*=\braket{l|M|l'}^* = \braket{l'|M|l}=M_{l'l}$, and hence is a valid observable.
If we have access to multiple copies of the states $\phi_l$ we can estimate the overlaps in the denominator. 
For small $L$ we can construct the matrix $M$ and compute its eigenvalues efficiently. Moreover, we can classically compute a unitary $U_M$ that diagonalizes $M$, i.e. $U_M^\dagger M U_M$ is diagonal, and find a quantum circuit that implements it on a quantum computer. Then the measurement of $M$ can be achieved by first applying $U_M$ and then measuring in the computational basis and recording the associated eigenvalue. Thus the circuit shown in Fig.~\ref{fig:lcs} can be used to prepare the weighted state $\ket{ \Phi}$.

Similarly, to the method for generating linear combinations of states via repeated applications of the Quantum State Polynomial algorithm, this algorithm breaks down if any one of the states $\{ \ket{ \phi_l} \}_{l=0}^L$ is outside the space spanned by the rest of the states. In the next two subsections we propose alternative methods that may be used in these cases.

\subsection{Incoherent post-processing method}\label{ap:postproc}

If our goal is simply to apply some unitary $V$ to $\ket{ \Phi}$ and make a measurement of $O$, we can achieve this by applying unitaries to $\{\ket{\phi_l}\}$ individually and computing matrix elements of $M$ in this new set. That is one can compute
\begin{align}
    \braket{ \Phi|V^\dagger O V| \Phi} &= \sum_{l,l'=0}^L \alpha_l \alpha_{l'}^* \braket{\phi_{l'}|V^\dagger O V|\phi_l} \, 
\end{align}
by computing each $\braket{\phi_{l'}|V^\dagger O V|\phi_l}$ term, reweighting it by $\alpha_l \alpha_{l'}^*$ and then summing together the outputs.

Computing the diagonal terms $\braket{\phi_{l}|V^\dagger O V|\phi_l}$ is straightforward. This is simply done by evolving $\ket{\phi_l}$ under $V$ and then measuring $O$. The off diagonal terms $\braket{\phi_{l'}|V^\dagger O V|\phi_l}$, for $l \neq l'$, can be computed using the Hadamard test if one has access to the controlled versions of the unitaries $W_l$ that prepare $\ket{\phi_l}$. To do so, we have to first expand the measurement operator as as a linear combination of unitaries, e.g. as $O = \sum_i r_i U_i$. Equipped with this linear combination we then have
\begin{align}
    \braket{\phi_{l'}|V^\dagger O V|\phi_l} &= \sum_i r_i \braket{0|W_{l'}^\dagger V^\dagger U_i VW_l|0} \, .
\end{align}
Thus $\braket{\phi_{l'}|V^\dagger O V|\phi_l} $ is a weighted sum of the expectation value of the unitaries $W_{l'}^\dagger V^\dagger U_i VW_l$ in the state $\ket{0}$, and each of these expectation values may be computed with a Hadamard test.
In applications where the gate sequence for implementing $W_l$ is known, a controlled version can be obtained with a constant factor overhead.

\subsection{Linear Combination of Unitaries (LCU) method}

Here we assume that we have access to the controlled versions of the unitaries $W_l$ that prepare $\ket{\phi_l}$. In order to prepare the linear combination of states in Eq.~\eqref{eq:LCS} we act on the $\ket{0}$ state with the LCU given by
\begin{align}
    \sum_{l=0}^{L} \alpha_l W_l \, .
\end{align}
The LCU method~\cite{childs2012hamiltonian} has the same restriction as the incoherent method in that it requires access to the controlled version of the unitaries $W_l$. The main difference is that the LCU method actually prepares the normalized state
\begin{align}
    \ket{\Phi_N} &= \frac{\ket{ \Phi}}{\Vert \ket{ \Phi}\Vert} = \frac{1}{\sqrt{\sum_{l l'} \alpha_l^* \alpha_{l'} \braket{\phi_l | \phi_{l'}}}}\sum_{l=0}^{L} \alpha_l \ket{\phi_l}
\end{align}
with probability $(\Vert \ket{ \Phi}\Vert/\Vert \alpha \Vert_1)^2$, where $\Vert \alpha \Vert_1=\sum_l \vert \alpha_l \vert$. This probability can be boosted by amplitude amplification. In addition, note that we need to know $\Vert \ket{ \Phi}\Vert$ in order to compute quantities of interest in terms of $\ket{ \Phi}$ using the actual state $\ket{\Phi_N}$. This can be done by computing the overlaps $\braket{\phi_l|\phi_{l'}}=\braket{0|W^\dagger_lW_{l'}|0}$ using the Hadamard test, since we assume access to the controlled-$W_l$'s.

\section{Linear state transformations via quantum teleportation}\label{app:teleportation}

\begin{figure}
    \centering
    \includegraphics[width=1\textwidth]{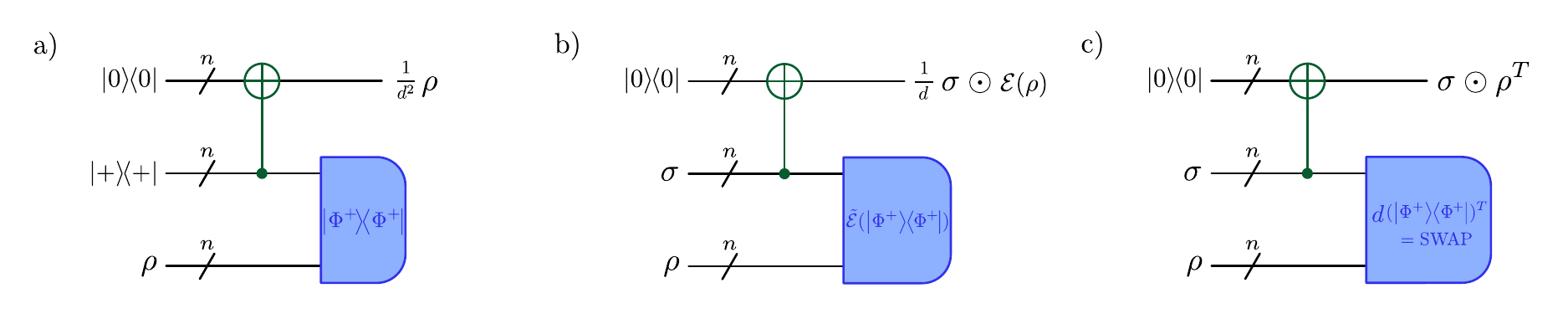}
    \caption{\textbf{Linear state transformations via teleportation.} a) We show the standard probabilistic circuit for quantum teleportation. With probability $1/d^2$ the state $\rho$ is teleported from the lower to upper register. Or, in the language of this paper, the output from the quantum instrument is the weighted state $\rho/d^2$. b) We show a generalized version of the teleportation circuit where i. instead of inputting the ancilla state $|+\rangle\langle+|$ one may input a generic state $\sigma$ and ii. instead of measuring the Bell state $|
    \Phi^+ \rangle \langle \Phi^+|$ one measures its mapped variant $ \tilde{\mathcal{E}}\left(|
    \Phi^+ \rangle \langle \Phi^+| \right)$. In this case the output state is $\frac{1}{d} \sigma \odot \mathcal{E}\left(|
    \Phi^+ \rangle \langle \Phi^+| \right)$. c) This generalized teleportation algorithm is equivalent to the generalized transpose algorithm (see Fig.~\ref{fig:gqt}) for the case where $  \tilde{\mathcal{E}}\left(|
    \Phi^+ \rangle \langle \Phi^+| \right) =  \mathcal{E} \left(|
    \Phi^+ \rangle \langle \Phi^+| \right) =d \left(|
    \Phi^+ \rangle \langle \Phi^+| \right)^T = \text{SWAP}$.}
    \label{fig:teleportation}
\end{figure}

In conventional quantum teleportation, shown in Fig.~\ref{fig:teleportation}a), a quantum state is teleported from some register $A$ to register $B$ as 
\begin{equation}
    \Tr_{AB} \left[ \left( \rho_{A} \otimes | \Phi^+ \rangle \langle \Phi^+ |_{BC} \right) ( | \Phi^+ \rangle \langle \Phi^+ |_{AB} \otimes \id_C ) \right] = \frac{1}{d^2} \rho_{C} \, 
\end{equation}
by preparing a Bell state $| \Phi^+ \rangle \langle \Phi^+ |_{BC}$ on registers $B$ and $C$ and then performing a Bell measurement $| \Phi^+ \rangle \langle \Phi^+ |_{AB}$ on registers $A$ and $B$.
In this section we show how linear, potentially non-positive, maps can be implemented on a quantum state via a generalization of the basic quantum teleportation algorithm. We further show how the generalized transpose algorithm can be understood within this framework.

The first possible generalization of the quantum teleportation involves changing the input state. Rather than preparing the initial ancillary qubits in a Bell state, the ancillary qubits are prepared in the state
\begin{equation}
    \rho_{\rm in}^{(\sigma)} = \text{CNOT} \left( \sigma \otimes |0\rangle\langle 0 | \right) \text{CNOT} \, 
\end{equation}
where $\sigma$ is an arbitrary quantum state.
We note that if $\sigma = | + \rangle \langle + |$ we get back the standard teleportation protocol with $\rho_{\rm in}^{(\sigma)} = |\Phi^+ \rangle \langle \Phi^+ |$. However, the effect of performing teleportation with arbitrary $\sigma$ is to reweigh the teleported state with the amplitudes of sigma. That is, the output teleported state is given by the quantum Hadamard product $\frac{1}{d} \sigma \odot |\psi\rangle \langle \psi |$. Thus this algorithm provides another means to implement the quantum Hadamard product of two states.

For the second generalization, instead of measuring the Bell state $| \Phi^+ \rangle \langle \Phi^+ |_{AB}$ one can measure the operator $\tilde{\mathcal{E}}_A (| \Phi^+ \rangle \langle \Phi^+ |_{AB})$ where we have introduced the map $\tilde{\mathcal{E}}_X = \sum_i J_X^{(i)} ( \, ... \, ) K_X^{(i)}$ with $J_X^{(i)} $ and $K_X^{(i)} $ generic operators. In this case, rather than teleporting the state $\frac{1}{d^2}\rho$ we teleport $\frac{1}{d^2} \mathcal{E}(\rho)$ where $\mathcal{E}_X(...) = \sum_i K_X^{(i)} \, ... \,  J_X^{(i)}$. To see why this works, note that
\begin{equation}
\begin{aligned}
         \frac{1}{d^2} \mathcal{E}_C(\rho_{C}) &= \Tr_{AB} \left[ \left( \mathcal{E}_A( \rho_{A} ) \otimes | \Phi^+ \rangle \langle \Phi^+ |_{BC} \right) \left( | \Phi^+ \rangle \langle \Phi^+ |_{AB} \otimes \id_C \right) \right] \\
        &= \sum_i \Tr_{AB} \left[ \left( K_A^{(i)} \rho_{A}  J_A^{(i)} \otimes | \Phi^+ \rangle \langle \Phi^+ |_{BC} \right) \left( | \Phi^+ \rangle \langle \Phi^+ |_{AB} \otimes \id_C \right) \right] \\
        &= \sum_i  \Tr_{AB} \left[ \left(  \rho_{A}  \otimes | \Phi^+ \rangle \langle \Phi^+ |_{BC} \right) \left(J_A^{(i)} | \Phi^+ \rangle \langle \Phi^+ |_{AB} K_A^{(i)} \otimes \id_C \right) \right] \\
        &= \Tr_{AB} \left[ \left(  \rho_{A} \otimes | \Phi^+ \rangle \langle \Phi^+ |_{BC} \right) \left( \tilde{\mathcal{E}}_A \left(| \Phi^+ \rangle \langle \Phi^+ |_{AB} \right) \otimes \id_C \right) \right] \, .
\end{aligned}
\end{equation}
That is, one can use this method to implement the transformation  $\rho_{A} \, \rightarrow \, \frac{1}{d^2} \mathcal{E}_C(\rho_{C})$. 

These two generalizations can be combined to implement the transformation 
\begin{equation}
  \frac{1}{d} \sigma \odot   \mathcal{E}( |
\psi \rangle \langle \psi | ) \, ,
\end{equation}
as shown in Fig.~\ref{fig:teleportation}b).

In the case of the generalized transpose operation we have $\mathcal{E}( \, ... \,  ) = \tilde{\mathcal{E}}( \, ... \,  ) = d ( ... )^T$ and thus the required measurement is $ d ( | \Phi^+ \rangle \langle \Phi^+ |_{AB} )^{T_A} = \text{SWAP}_{AB}$.~\footnote{To see this note that the transpose operation on a single qubit can be implemented as $( ... )^T = |0\rangle \langle 0 | \, ... \, |0\rangle \langle 0 | +  |1\rangle \langle 1 | \, ... \, |1 \rangle \langle 1 | +  |1\rangle \langle 0 | \, ... \, |1\rangle \langle 0 | +  |0\rangle \langle 1 | \, ... \, |0\rangle \langle 1 |$. The generalization to multiple qubits can be seen by considering the tensor product of multiple transpose maps.}  Thus, as is apparent from comparing Figs.~\ref{fig:teleportation} and Fig.~\ref{fig:gqt}, we see that the generalized transpose algorithm, is effectively a special case of this more general method to implement a mapping $ \frac{1}{d} \sigma \odot \mathcal{E}( |
\psi \rangle \langle \psi | )$ via quantum teleportation.

\medskip

A natural question to ask is what class of mappings can be implemented via this method. If we require the measurement operator to be a Hermitian observable then we require that $\tilde{\mathcal{E}}_A (| \Phi^+ \rangle \langle \Phi^+ |_{AB}) = \tilde{\mathcal{E}}_A (| \Phi^+ \rangle \langle \Phi^+ |_{AB})^\dagger$, or equivalently that 
$\sum_i J_A^{(i)} (| \Phi^+ \rangle \langle \Phi^+ |_{AB}) K_A^{(i)} = \sum_i  K_A^{(i)\dagger} (| \Phi^+ \rangle \langle \Phi^+ |_{AB}) J_A^{(i)\dagger} $. 
These identities constrain the allowed operators for $J_A^{(i)}$ and $K_A^{(i)}$ and so the maps $\mathcal{E}$ that may be implemented.

However, one could more generally allow the measurement operation to be a normal operator or a probabilistic combination of normal operators. For example, for the case of a single qubit one can implement a transformation that swaps the order of the two rows of the density operator, i.e.,
\begin{equation}
    \begin{bmatrix}
        |\alpha|^2 & \alpha \beta^* \\
        \alpha^* \beta & |\beta|^2
    \end{bmatrix} \rightarrow 
       \begin{bmatrix}
       \alpha^* \beta & |\beta|^2 \\
        |\alpha|^2 & \alpha \beta^* 
    \end{bmatrix}
\end{equation}
by measuring $|\psi^- \rangle \langle \phi^+ |$. This operator is non-normal but can be ``measured'' by measuring $|\psi^- \rangle \langle \phi^+ | + |\phi^+ \rangle \langle \psi^- |$ and $ i \left( |\psi^- \rangle \langle \phi^+ | - |\phi^+ \rangle \langle \psi^- | \right)$ with equal probabilities and multiplying the output of the second measurement by a factor of $-i$.

\section{Sampling complexity analysis}\label{ap:Errors}

\subsection{General analysis for weighted states}

In the approach outlined in Section~\ref{sec:ES} the expectation value of an operator $O$ in a weighted state $\tilde{\tau}$ is obtained from the expectation value of the operator $O\otimes M$ in a state $\rho^\text{out}$ --- see Eq.~\eqref{eq:map}. In this appendix we analyze the sampling complexity of estimating expectation values using this approach in general before focusing on specific applications in the following sections. 
Unlike in the main text where $M$ was Hermitian, here we allow it to be an arbitrary normal operator.
The total operator whose expectation value we need to estimate is given by
\begin{align}
    T =O \otimes M  \otimes \mathbb{I}
\end{align}
We estimate $\langle T \rangle$ with $\widehat{\mathcal{T}}$ using 
\begin{align}
    \widehat{\mathcal{T}} &= \frac{1}{s} \sum_{i=1}^{s} \nu_{t(i)} \;.
\end{align}
Here, $s$ is the number of circuit evaluations and $\nu_{t(i)}$ is the outcome of $i$'th measurement i.e. the $t(i)$'th eigenvalue of $T$.  
Since $T$ is a normal operator $\widehat{\mathcal{T}}$ is a complex random number. 
This estimator is unbiased since it relies on the standard method for estimating a quantum expectation value. 
The variance is defined as
\begin{align}
    \text{Var}(\widehat{\mathcal T}) &= E[|\widehat{\mathcal T}|^2]-|E[\widehat{\mathcal T}]|^2 \\
    &= \frac{1}{s} \left( \Tr[\rho^\text{out}(OO^\dagger\otimes  M M^\dagger \otimes \mathbb{I})]-|\Tr[\tau O]|^2 \right) \, .
\end{align}

In Appendix.~\ref{ap:beyondnormal} we explained how to express a quantum instrument with a non-normal measurement operator $\overline{M}=\sum_k c_k N_k$ in terms of an instrument with a normal measurement operator $M$. In this case the above analysis can be repeated to yield
\begin{align}
    \text{Var}(\widehat{\mathcal T}) &= \frac{1}{s} \left(\sum_k c_k \Tr[\bar\rho^\text{out}(OO^\dagger\otimes  N_k N_k^\dagger \otimes \mathbb{I})]-|\Tr[\tau O]|^2 \right)
\end{align}
Ideally one wants to minimize the number of shots $s$, i.e. circuit evaluations, needed to achieve a desired variance for the estimator. 
This can be done, in theory, by optimizing over all quantum instruments \{$\mathcal{I}(\sigma,U,M)\}$ that implement the desired transformation. 
However, it is an open question how to do this in practice. 
In addition, there may be other considerations than sampling complexity. For instance, in order to measure a normal operator we first need to apply the diagonalizing unitary, which might be hard to compute classically or hard to implement on a quantum computer. Such considerations should also be taken into account when formulating the optimization task, especially when using noisy quantum devices.

\subsection{Proof of operator independent bound on variance}
To compare the sampling complexity associated with different quantum instruments implementing the same transformation and provide an observable independent analysis of the complexity of our algorithms, it is advantageous to derive an operator independent bound on the variance. Here we provide proofs of our operator independent bounds on the variance quoted in the main text. We start by providing a proof~\cite{stackexchangebound}   for the following bound on the joint variance
\begin{equation}\label{eq:RawErrorBoundAp}
       \sigma_{XY}^2 \leq 2 \sigma_{X}^2 || Y ||_\infty^2 + 2 |\langle X \rangle|^2 \sigma_{Y}^2 \,  
\end{equation}
where $X$ and $Y$ are complex random variables.
To derive this we consider two complex random variables $A =( X - \langle X \rangle) Y $ and $B = \langle X \rangle Y $, and note that 
\begin{align}
    \text{Var}( X Y ) &= \text{Var}( A + B ) \\
      &\leq 2 \text{Var}( A ) +  2 \text{Var}( B  ) \\ 
      &= 2 \text{Var}( ( X - \langle X \rangle) Y ) + 2 |\langle X \rangle|^2 \text{Var}(Y )  \, .
\end{align}
We then note that,
\begin{align}
    \text{Var}( ( X - \langle X \rangle) Y ) &\leq \langle | X - \langle X \rangle) Y|^2 \rangle \\
     &\leq  \text{Var}(X) ||| Y ||_\infty|^2 \, ,
\end{align}
and thus we obtain Eq.~\eqref{eq:RawErrorBoundAp}. Applying this to evaluating $\text{Var}(\widehat{\mathcal{O}})$, and assuming $O$ is Hermitian, we have that
\begin{equation}
  \text{Var}(\widehat{\mathcal{O}})  \leq \frac{2}{s } \left(  \sigma_{M}^2 || O ||_\infty^2 + |\langle M \rangle|^2 \sigma_{O}^2 \right) \, .
\end{equation}
Assuming that the largest eigenvalue of $O$ is bounded by 1 we have that $ || O ||_\infty^2 \leq 1$ and $\sigma_{O}^2 \leq 1$ and so 
\begin{equation}\label{eq:ErrorBoundApOld}
  \text{Var}(\widehat{\mathcal{O}})  \leq \frac{2}{s} \left(  \sigma_{M}^2 + |\langle M \rangle|^2  \right) = \frac{2 \langle |M|^2 \rangle }{s} \, .
\end{equation}

\zo{Alternatively, we can bound $\text{Var}(\widehat{\mathcal{O}}) $ as follows
\begin{equation}
\begin{aligned}\label{eq:ErrorBoundAp}
    \text{Var}(\widehat{\mathcal{O}})  &= E[|\widehat{\mathcal O}|^2]-|E[\widehat{\mathcal O}]|^2  \\    &= \frac{1}{s} \left( \Tr[\rho^\text{out}(OO^\dagger\otimes  M M^\dagger \otimes \mathbb{I})]-|\Tr[\tau O]|^2 \right) \\ &\leq \frac{1}{s} \Tr[\rho^\text{out}(OO^\dagger\otimes  M M^\dagger \otimes \mathbb{I})] \\  &= \frac{1}{s} \sum_{ij} p(o_i, m_j) |o_i |^2 |m_j |^2  \\  
     &\leq \frac{|| O ||_\infty^2}{s} \sum_{j}
    p( m_j) |m_j|^2  \\
     &= \frac{|| O ||_\infty^2 \langle |M|^2 \rangle}{s} \\
     &\leq \frac{|| O ||_\infty^2 ||M ||_\infty^2}{s} 
\end{aligned}
\end{equation}
In the limit that $ || O ||_\infty^2 \leq 1$ this bound is tighter than Eq.~\eqref{eq:ErrorBoundApOld} by a factor of 2.}
In the following sections we evaluate the variance $\text{Var}(O)$ and its bound,
\begin{equation}
    \mathcal{B}_{\rm Var} :=  \frac{\langle |M|^2 \rangle }{s} \, ,
\end{equation}
for the primitive nonlinear subroutines that we have introduced in this manuscript.

\zo{\subsection{Hoeffding's inequality analysis}

Hoeffding's upper bound on the probability that the sum of bounded independent random variables deviates from its expected value provides an alternative means of analysing the convergence of our weighted state algorithms. Specifically, Hoeffding's inequality entails that 
\begin{equation}
    P \left( \vert\hat{\mathcal{T}} - \langle T \rangle\vert \geq \epsilon \right) \leq 2e^{\frac{-2 N \epsilon^2}{(\mu_{\rm max} - \mu_{\rm min})^2}} 
\end{equation}
where $\mu_{\rm max}$ and $ \mu_{\rm min}$ are the maximum and minimum eigenvalues of $T = O \otimes M $ respectively.

We have that $|\mu_{\rm max} - \mu_{\rm min}| \leq 2\|O\otimes M\|_\infty = 2\|O\|_\infty \|M\|_\infty$. It thus follows that the probability that the computed value of the target observable $O$ deviates from the true average $\langle O \rangle_\tau$ in weighted state $\tau$ can be bounded as 
\begin{equation}
    P \left( \vert\hat{\mathcal{T}} - \langle O \rangle_\tau\vert \geq \epsilon \right) \leq 2 e^{\frac{- N \epsilon^2}{2|| O ||_\infty^2 || M ||_\infty^2}}  \, .
\end{equation}
Or, turning it around, to ensure that the deviation from the true mean is greater than $\epsilon$ occurs with probability at most $\delta$, i.e. $P \left( \vert\hat{\mathcal{T}} - \langle O\vert \rangle_\tau \geq \epsilon \right) \leq \delta$, the number of shots required is
\begin{equation}
    N \geq \frac{2||O||_\infty^2 || M||_\infty^2 \log(\frac{2}{\delta})}{ \epsilon^2} \, .
\end{equation}
Thus we see find the same quadratic dependence on $|| O ||_\infty^2$ and $|| M ||_\infty^2$ as in the original analysis in Eq.~\eqref{eq:ErrorBoundAp}.
}

\subsection{Quantum state polynomial}

For the case of the quantum state polynomial algorithm the variance of the estimator evaluates to 
\begin{equation}
\begin{aligned}
           \text{Var}(\widehat{\mathcal{O}})  = \frac{1}{ s}  \bigg( &\sigma_{00} [M^\dagger M]_{00}  \Tr[\rho_0 O^2 ] +  \sigma_{11} [M^\dagger M]_{11}  \Tr[\rho_1 O^2  ] \\ &+ \sigma_{01} [M^\dagger M]_{10} \Tr[\rho_0 \rho_1 O^2 ] +  \sigma_{10} [M^\dagger M]_{01} \Tr[ \rho_1 \rho_0 O^2] - |\Tr[ \tau O ]|^2 \bigg) \, .
\end{aligned}
\end{equation}
Since we have $\alpha = \sigma \odot M^T \odot \gamma^{\rm in}$, it follows that 
\begin{equation}
\begin{aligned}
       \text{Var}(\widehat{\mathcal{O}}) = \frac{1}{s} \bigg( &\sigma_{00} \left( \frac{|\alpha_{00}|^2}{|\sigma_{00}|^2} + \frac{|\alpha_{10}|^2}{|\sigma_{10}|^2} \right) \Tr[\rho_1] \Tr[\rho_0 O^2 ] +  \sigma_{11}  \left( \frac{|\alpha_{11}|^2}{|\sigma_{11}|^2} + \frac{|\alpha_{01}|^2}{|\sigma_{10}|^2}\right) \Tr[\rho_0] \Tr[\rho_1 O^2  ] \\ +  &  \left( \frac{\alpha_{00}}{\sigma_{00}} \frac{\alpha_{01}^*}{\sigma_{01}^*} + \frac{\alpha_{11}^*}{\sigma_{11}^*} \frac{\alpha_{10}}{\sigma_{10}}\right) \sigma_{01} \Tr[\rho_0 \rho_1 O^2 ] +   \left( \frac{\alpha_{00}^*}{\sigma_{00}^*} \frac{\alpha_{01}}{\sigma_{01}} + \frac{\alpha_{11}}{\sigma_{11}} \frac{\alpha_{10}^*}{\sigma_{10}^*}\right) \sigma_{10} \Tr[ \rho_1 \rho_0 O^2]  - |\Tr[ \tau O ]|^2  \bigg) \, .
\end{aligned}
\end{equation}
We note that this expression is real since the off diagonal elements are complex conjugate of one another. 
We can use the operator independent bound, Eq.~\eqref{eq:ErrorBoundAp}, to bound this variance as
\begin{equation}
\begin{aligned}
      \text{Var}(\widehat{\mathcal{O}})  \leq \frac{2 }{s} \Bigg( & \left(\sigma_{00} \left( \frac{\alpha_{00}^2}{\sigma_{00}^2} + \frac{\alpha_{01}\alpha_{10}}{\sigma_{01}\sigma_{10}} \right) +  \sigma_{11}  \left( \frac{\alpha_{11}^2}{\sigma_{11}^2} + \frac{\alpha_{01}\alpha_{10}}{\sigma_{01}\sigma_{10}} \right) \right) \Tr[\rho_0] \Tr[\rho_1 ] \\ +  &  \left( \frac{\alpha_{00}}{\sigma_{00}} + \frac{\alpha_{11}}{\sigma_{11}} \right) \left( \alpha_{01} +  \alpha_{10} \right) \Tr[ \rho_0 \rho_1]  \Bigg) \, .
\end{aligned}
\end{equation}

\subsection{Linear combination of states via Quantum State Polynomial algorithm}

For the special case of a linear combination of states we have $\alpha_{00} \rightarrow |\alpha_0|^2/\langle \psi_0 | \psi_0 \rangle$, $\alpha_{11} \rightarrow |\alpha_1|^2/\langle \psi_1 | \psi_1 \rangle$, $\alpha_{01} \rightarrow \alpha_0 \alpha_1^* / \langle \psi_0 | \psi_1 \rangle$, $\alpha_{10} \rightarrow \alpha_1 \alpha_0^* / \langle \psi_1 | \psi_0 \rangle$ and $\sigma_{00} \rightarrow |\beta_0|^2$, $\sigma_{11} \rightarrow |\beta_1|^2$, $\sigma_{01} \rightarrow \beta_0 \beta_1^*$, $\sigma_{10} \rightarrow \beta_1 \beta_0^*$. Therefore in this case we have 
\begin{equation}
\begin{aligned}\label{eq:VarLinComboPair}
       \text{Var}(\widehat{\mathcal{O}})  = \frac{1}{s } \Bigg( &|\beta_0|^2 N_1 \left( \frac{|\alpha_0|^4}{|\beta_0|^4 N_0} + \frac{|\alpha_0|^2|\alpha_{1}|^2}{|\beta_{0}|^2|\beta_{1}|^2 r}  \right) \Tr[\rho_0 O^2 ] +  |\beta_1|^2 N_0 \left( \frac{|\alpha_1|^4}{|\beta_1|^4 N_1}  + \frac{|\alpha_0|^2|\alpha_{1}|^2}{|\beta_{0}|^2|\beta_{1}|^2 r} \right)  \Tr[\rho_1 O^2  ] \\ +  &  2\left( \frac{|\alpha_0|^2}{|\beta_0|^2 \sqrt{N_0}}  + \frac{|\alpha_1|^2}{|\beta_1|^2 \sqrt{N_1}}  \right) \Re \left( \alpha_{0} \alpha_{1}^* \langle \psi_0 | O^2 | \psi_1 \rangle \right)- \Tr[ \tau O ]^2  \Bigg) \, \\
       = \frac{1}{s } \Bigg( & |\alpha_0|^2 N_1 \left( \frac{|\alpha_0|^2}{|\beta_0|^2 N_0} + \frac{\alpha_{1}|^2}{|\beta_{1}|^2 r}  \right) \Tr[\rho_0 O^2 ] +  |\alpha_1|^2 N_0 \left( \frac{|\alpha_1|^2}{|\beta_1|^2 N_1}  + \frac{|\alpha_0|^2}{|\beta_{0}|^2 r} \right)  \Tr[\rho_1 O^2  ] \\ +  &  2\left( \frac{|\alpha_0|^2}{|\beta_0|^2 \sqrt{N_0}}  + \frac{|\alpha_1|^2}{|\beta_1|^2 \sqrt{N_1}}  \right) \Re \left( \alpha_{0} \alpha_{1}^* \langle \psi_0 | O^2 | \psi_1 \rangle \right)- \Tr[ \tau O ]^2 \Bigg) \, , 
\end{aligned}
\end{equation}
where $N_0 = | \langle \psi_0 | \psi_0 \rangle |^2$,  $N_1 = | \langle \psi_1 | \psi_1 \rangle |^2$, and $r = |\langle \psi_0 | \psi_1 \rangle |^2$.

With the weighted state method for generating linear combinations of states there is some freedom in how the final measurement operator is chosen depending on the state of the ancilla qubit. To assess the optimum strategy, i.e. the optimum ancilla state and measurement pair that minimizes the sampling complexity, it is advantageous to use the operator independent bound on the variance Eq~\eqref{eq:ErrorBoundAp}. 
It follows from Eq~\eqref{eq:ErrorBoundAp} and Eq.~\eqref{eq:VarLinComboPair} that
\begin{equation}
    \langle M^2 \otimes \mathbb{I} \rangle = \,  |\alpha_0|^2 \left(\frac{|\alpha_0|^2}{|\beta_0|^2}  + \frac{|\alpha_1|^2}{|\beta_1|^2r} \right)  + |\alpha_1|^2 \left( \frac{|\alpha_0|^2}{|\beta_0|^2 r} + \frac{|\alpha_1|^2}{|\beta_1|^2} \right) \, .
\end{equation}
Setting $|\alpha_0|^2 = p$ and $|\beta_0|^2 = q$ this can be rewritten as 
\begin{equation}
    \langle M^2 \otimes \mathbb{I} \rangle = \,  p \left( \frac{p}{q}  + \frac{(1-p)}{(1-q)r} \right)  + (1-p) \left( \frac{p}{q r} + \frac{(1-p)}{(1-q)} \right) = f(p,q,r) \, .
\end{equation}
Given $0 \leq p \leq 1$ and $0 \leq r \leq 1$, $f(p,q,r)$ is minimized when
\begin{equation}\label{eq:optBeta}
    q_{\rm opt} (p, r) = \frac{2 p - 2 p^2 + 2 p^2 r + 2 r \sqrt{\frac{(-p + p^2) (-p + p^2 - r + 2 p r - 2 p^2 r - p r^2 + p^2 r^2)}{ r^2}}}{2 \left(2 p - 2 p^2 + r - 2 p r + 2 p^2 r + 
   2 r \sqrt{\frac{(-p + p^2) (-p + p^2 - r + 2 p r - 2 p^2 r - p r^2 + 
       p^2 r^2)}{r^2}}\right)}
\end{equation}
This expression thus gives us a way of picking the ancilla state specified by $\beta_0$ given a target state specified by $\alpha_0$ and $r = | \langle \psi_0 | \psi_1 \rangle |^2$. In Fig.~\ref{fig:optstrat} we use $\eqref{eq:optBeta}$ to plot the optimum $|\beta_0|^2$ as a function of $|\alpha_0|^2$ and $r$. In the limit that $r \rightarrow 0$, we have that $q_{\rm opt} = \rightarrow 1/2$, that is the optimum strategy is picking $|\beta_0| \approx 1/\sqrt{2}$. Conversely, in the limit that $r \rightarrow \infty$ we have $  |\beta_0|^2 = q_{\rm opt} \rightarrow \frac{p}{p + \sqrt{(1 - p)p}}$.

\subsection{Linear combination of states via incoherent post-processing method}

Suppose instead we compute the expectation value of an observable $O$ with respect to the state $\ket{\psi}$ using purely classical post-processing. That is, as discussed in Appendix~\ref{ap:postproc}, we compute 
\begin{equation}\label{eq:PostPro}
\begin{aligned}
        \bra{\psi} O \ket{\psi} &= |\alpha_0|^2 \bra{\psi_0} O \ket{\psi_0}  + |\alpha_1|^2 \bra{\psi_1} O \ket{\psi_1} + \alpha_0^*\alpha_1  \bra{\psi_0} O \ket{\psi_1}  + \alpha_0 \alpha_1^*  \bra{\psi_1} O \ket{\psi_0} \\ 
        &= |\alpha_0|^2 \bra{\psi_0} O \ket{\psi_0}  + |\alpha_1|^2 \bra{\psi_1} O \ket{\psi_1} + 2\Re(\alpha_0^*\alpha_1)  \Re(\bra{\psi_0} O \ket{\psi_1}) - 2\Im(\alpha_0^* \alpha_1) \Im(\bra{\psi_0} O \ket{\psi_1})
\end{aligned}
\end{equation}
by computing the terms $\bra{\psi_0} O \ket{\psi_0}$, $\bra{\psi_1} O \ket{\psi_1}$, $\Re(\bra{\psi_0} O \ket{\psi_1})$ and $\Im(\bra{\psi_0} O \ket{\psi_1})$ separately. To measure the off-diagonal terms $\Re(\bra{\psi_0} O \ket{\psi_1})$ and $\Im(\bra{\psi_0} O \ket{\psi_1})$ we can expand the measurement operator $O$ as a sum of unitaries\footnote{Without loss of generality we here assume that the $\eta_i$ prefactors are real and positive, $\eta_i > 1$, by absorbing any phases into the corresponding unitary $U_i$.}, i.e. $
O = \sum_i \eta_i U_i$, and then use pairs of Hadamard tests to obtain the real and imaginary parts of the overlap terms $\bra{\psi_0} U_i \ket{\psi_1}$. That is, in total one evaluates each of the terms in
\begin{equation}\label{eq:PostProExpl}
\begin{aligned}
        \bra{\psi} O \ket{\psi} &= |\alpha_0|^2 \bra{\psi_0} O \ket{\psi_0}  + |\alpha_1|^2 \bra{\psi_1} O \ket{\psi_1} + 2 \sum_i \eta_i \left( \Re(\alpha_0^*\alpha_1)  \Re(\bra{\psi_0} U_i \ket{\psi_1}) - \Im(\alpha_0^* \alpha_1) \Im(\bra{\psi_0} U_i \ket{\psi_1}) \right) \, .
\end{aligned}
\end{equation}

Let use denote the estimators of the terms $\bra{\psi_0} O \ket{\psi_0}$, $\bra{\psi_1} O \ket{\psi_1}$, $ \Re(\bra{\psi_0} U_i \ket{\psi_1})$, and $\Im(\bra{\psi_1} U_i \ket{\psi_0})$ in Eq.~\eqref{eq:PostProExpl} by $\widehat{\mathcal{P}}_{0}$, $\widehat{\mathcal{P}}_{1}$, $ \widehat{\mathcal{P}}_{2,i}$ and $ \widehat{\mathcal{P}}_{3,i}$ respectively.
Following the approach of Ref.~\cite{arrasmith2020operator} we will suppose that our total shot quota is divided between the different circuits in proportion to the prefactor in the sum. 
Under this approach, for a random variable of the form $Q = \sum_i \mu_{i} Q_{i}$, we use $s_{i} \propto |\mu_{i}|$ shots to get the estimator $\widehat{\mathcal{Q}}_{i}$, resulting in a variance of
\begin{equation}
    \text{Var}(\widehat{\mathcal{Q}})= \sum_{i} \frac{|\mu_{i}|^2 \text{Var}(\widehat{\mathcal{Q}}_{i}) }{s_{i}} =\sum_{i} \frac{|\mu_{i}|^2 \text{Var}(\widehat{\mathcal{Q}}_{i}) }{\lfloor|\mu_{i}| \tilde{s}\rfloor} \approx \sum_{i} \frac{|\mu_{i}| \text{Var}(\widehat{\mathcal{Q}}_{i}) }{ \tilde{s}} \, .
\end{equation}
Here the normalisation term $\tilde{s}$ is chosen such that $\sum_i s_i = \sum_i \lfloor   |\mu_i| \tilde{s}\rfloor = s$ and the approximation at the last equality gets better with the number of shots. 
The variance for the incoherent post-processing method thus takes the form 
\begin{equation}
\begin{aligned}
    \text{Var}(\widehat{\mathcal{O}}) \approx  \frac{1}{\tilde{s}} \bigg( &|\alpha_0|^2 \text{Var}(\widehat{\mathcal{P}}_0) + |\alpha_1|^2 \text{Var}(\widehat{\mathcal{P}}_1) +  \\ &2 | \Re(\alpha_0^*\alpha_1)| \sum_i  \eta_i \text{Var}(\widehat{\mathcal{P}}_{2,i}) + 2 |\Im(\alpha_0^*\alpha_1)| \sum_i \eta_i \text{Var}(\widehat{\mathcal{P}}_{3,i})  \bigg) \, ,
\end{aligned}
\end{equation}
where $\tilde{s} \approx s \left(|\alpha_0|^2 + |\alpha_1|^2 + 2\left(|\Re(\alpha_0^*\alpha_1)| + |\Im(\alpha_0^*\alpha_1)| \right) \sum_i r_i\right)^{-1}$.

Since the Hadamard test only requires measuring a Pauli operator, the errors for computing the overlap terms take the form 
\begin{align}
    \text{Var}(P_{2i})  = 1 - \Re(\bra{\psi_0} U_i \ket{\psi_1})^2 \\ 
    \text{Var}(P_{3i})  = 1 - \Im(\bra{\psi_0} U_i \ket{\psi_1})^2  \, .
\end{align}
Thus the total error takes the form 
\begin{equation}
\begin{aligned}
    \text{Var}(\widehat{\mathcal{O}}) =  \frac{1}{\tilde{s}} \bigg(& |\alpha_0|^2 \left(\bra{\psi_0} O^2 \ket{\psi_0}  - \bra{\psi_0} O \ket{\psi_0}^2 \right) \\ &+ |\alpha_1|^2 \left(\bra{\psi_1} O^2 \ket{\psi_1}  - \bra{\psi_1} O \ket{\psi_1}^2 \right) \\ &+  2\Re(\alpha_0^*\alpha_1) \sum_i  r_i \left(1 - \Re(\bra{\psi_0} U_i \ket{\psi_1})^2 \right) \\ 
    &+ 2\Im(\alpha_0^*\alpha_1) \sum_i r_i \left(1 - \Im(\bra{\psi_0} U_i \ket{\psi_1})^2 \right) \bigg)\, .
\end{aligned}
\end{equation}

\begin{figure}
    \centering
    \includegraphics[width=0.95\textwidth]{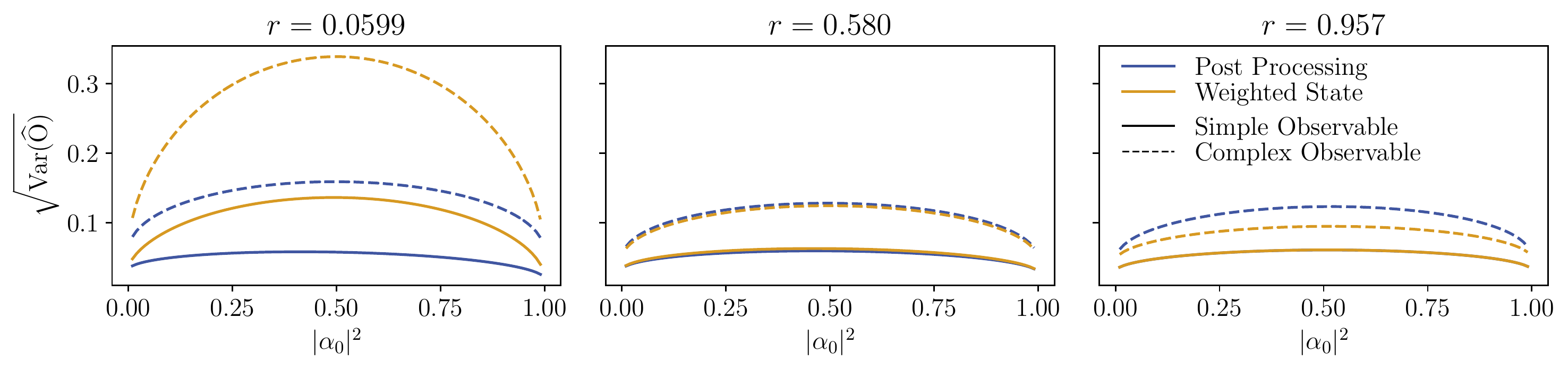}
    \caption{\textbf{Comparison of weighted state and post-processing methods.} Here we plot the standard deviation of the estimator, $\sqrt{\text{Var}(\widehat{\mathcal{O}})}$, for the weighted state method (yellow) and the post-processing method (blue) for implementing the linear combination of two $n=4$ qubit states as a function of $\alpha_0$. In both cases we suppose that a total of $s = 100$ shots are used. We plot the results for pairs of randomly generated states with a small overlap ($r = 0.067$, left), median overlap ($r = 0.58$, middle) and maximum overlap ($r = 0.95$, right). We compare the case of measuring a simple measurement (solid) composed of a single Pauli term and a complex measurement (dashed) composed of $n^2$ Pauli terms.}
    \label{fig:CompLinearCombo}
\end{figure}

In Fig.~\ref{fig:CompLinearCombo} we compare the convergence of the weighted state and post-processing methods for taking the linear combination of states. We find that with the exception of taking the linear combination of nearly orthogonal states the convergence is comparable. It is unsurprising that the convergence for nearly orthogonal states is poor for the weighted state method since, as discussed in the main text, this method cannot be applied to perfectly orthogonal states. In contrast for observables composed of a sum of many unitaries (i.e. `complex' observables in Fig.~\ref{fig:CompLinearCombo}) the weighted state approach has smaller variance than the post-processing method because a large number of distinct circuits need to be run for the latter.

\subsection{Powers of states via the generalized quantum transpose algorithm}

The Generalized Quantum Transpose algorithm may be used to prepare the Quantum Hadamard Product of a pair of states $\rho^{(0)}$ and $\rho^{(1)}$ if one knows how to prepare the transpose of either $\rho^{(0)}$ or $\rho^{(1)}$. Moreover, the procedure for preparing the transpose of a state may be straightforwardly computed if the procedure for preparing the original state is known in some detail. Thus GQT, under the right circumstances, provides an alternative means of implementing the QHP and so powers of states. In this section, we include a sampling analysis of this alternative means of implementing powers of states and compare it with the standard method of QHP.

When the GQT circuit is repeatedly applied to implement the power of a state, the variance is given by
\begin{equation}
    \text{Var}_{GQT}(\widehat{\mathcal{O}}) = \Tr[D(|\psi\rangle \langle \psi | )O^2] - \left( \bra{\psi^k} O \ket{\psi^k} \right)^2
\end{equation}
for any power $k$. 
Comparing this to the variance $\text{Var}_{\rm QHP}[\hat{\mathcal{O}}]$ obtained for the QHP algorithm we have 
\begin{equation}
\begin{aligned}\label{eq:diffvar}
        D &:= \text{Var}_{\rm QHP}[\hat{\mathcal{O}}] - \text{Var}_{\rm GQT}[\hat{\mathcal{O}}]  = \langle \psi^k | O^2 | \psi^k \rangle - \text{Tr}(D( | \psi \rangle \langle \psi |)O^2)  \\ 
        &= \sum_{ij} (\psi_j^*)^k (O^2)_{ji} \psi_i^k - \sum_i |\psi_i|^{2} (O^2)_{ii} \, .
\end{aligned} 
\end{equation}
To get a handle on this quantity it is helpful to first consider the case where $O^2 = 1$, as is the case, for example if $O$ is a Pauli operator. In this case, we have 
\begin{equation}
    D =  \langle \psi^k | \psi^k \rangle - 1 \, .
\end{equation}
Since $ \langle \psi^k | \psi^k \rangle \leq 1$ it follows that $\text{Var}_{\rm QHP}[\hat{\mathcal{O}}] \leq  \text{Var}_{\rm GQT}[\hat{\mathcal{O}}]$, i.e. the convergence of QHP is better than that of GQT. 

It is also straightforward to show that when the input states are pure single qubit states we also are guaranteed that $\text{Var}_{\rm QHP}[\hat{\mathcal{O}}] \leq  \text{Var}_{\rm GQT}[\hat{\mathcal{O}}]$. To see this, let us suppose $| \psi \rangle = \sqrt{ 1 - \epsilon } |0\rangle + \sqrt{\epsilon} |1 \rangle $, where we drop a relative phase between $|0\rangle$ and $|1 \rangle$ without loss of generality since this phase does not contribute to the variance. For the case of a single application $k=2$ we have
\begin{equation}
\begin{aligned}
    D &= \sum_{ij} (\psi_j^*)^2 (O^2)_{ji} \psi_i^2 - \sum_i |\psi_i|^{2} (O^2)_{ii} \\ 
    &= (1 - \epsilon)^2(O^2)_{00} + \epsilon^2 (O^2)_{11} + \epsilon (1 - \epsilon)  ((O^2)_{01} + (O^2)_{10}) - (1 - \epsilon)(O^2)_{00} - \epsilon (O^2)_{11} \\
    &= - \epsilon (1 - \epsilon) \left( (O^2)_{00} + (O^2)_{11} - (O^2)_{01} - (O^2)_{10} \right) \, . 
\end{aligned}
\end{equation}
Without loss of generality we can write 
$O = \alpha_I \mathbb{I} + \alpha_x X + \alpha_y Y + \alpha_z Z$ such that $O^2 = ( \alpha_I^2 + \alpha_x^2 + \alpha_y^2 + \alpha_z^2)\mathbb{I} + \alpha_I \alpha_x X + \alpha_I \alpha_y Y + \alpha_I \alpha_z Z$. Thus we end up with  
\begin{equation}
\begin{aligned}
        D &= - \epsilon  (1 - \epsilon) \left( \alpha_I^2 + \alpha_x^2 + \alpha_y^2 + \alpha_z^2   -   2 \alpha_I \alpha_x  \right)  \\ 
        &= - \epsilon  (1 - \epsilon) \left( \left( \alpha_I + \alpha_x \right)^2 + \alpha_y^2 + \alpha_z^2  \right)  \leq 0 \, .
\end{aligned}
\end{equation}
Thus the variance of QHP is less than that of GQT in the case that $k = 2$. It is manifestly clear from Eq.~\eqref{eq:diffvar} that increasing $k$ for a given $O$ and $|\phi\rangle$, decreases $D$ and hence $D$ is negative for all $k$ for the case of single qubit states.

\begin{figure*}[t!]
    \centering
    \includegraphics[width=0.48\textwidth]{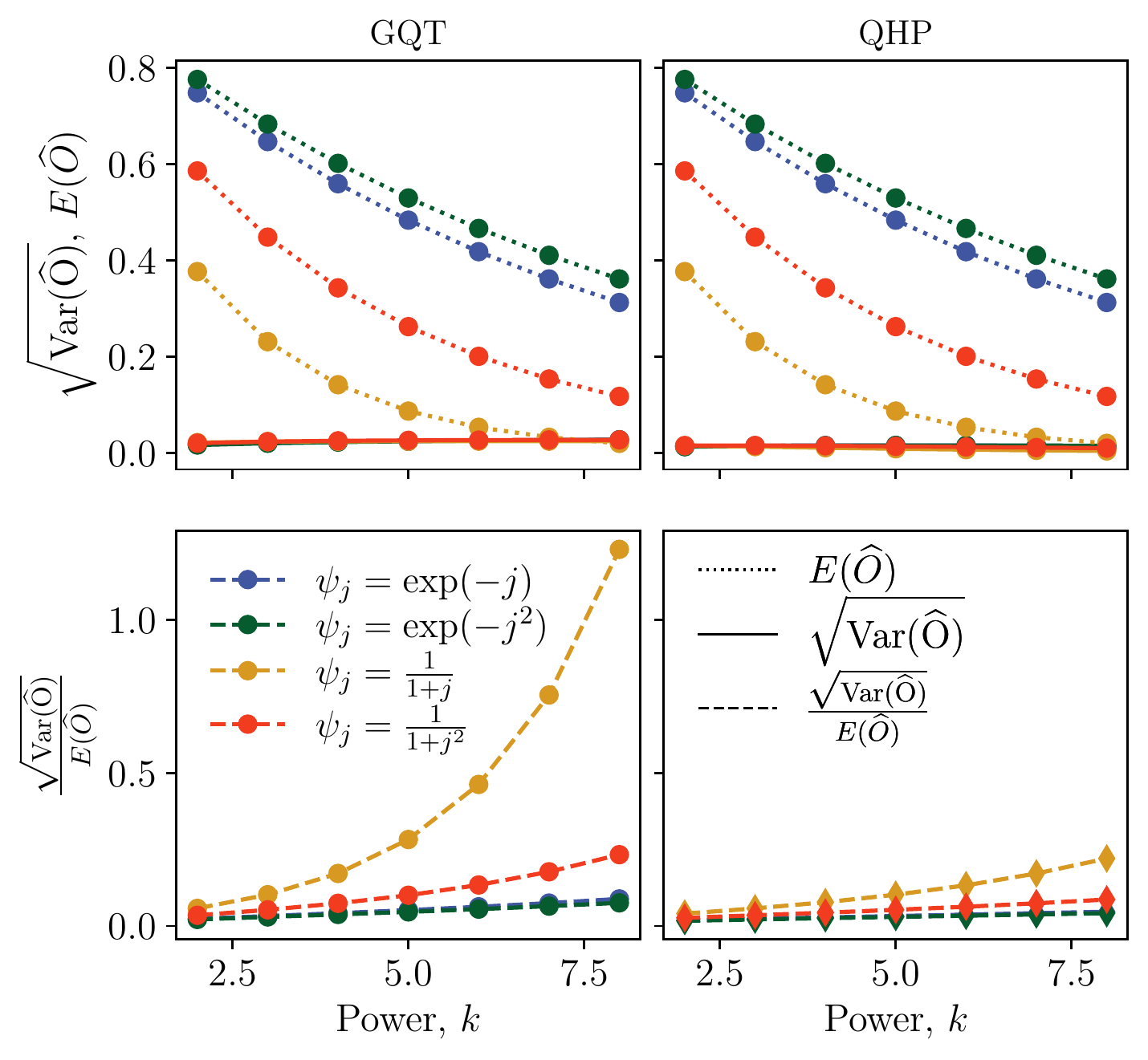}
    \caption{\textbf{ Comparing Sampling Error for Powers of States via QHP and GQT.} Here we plot the standard deviation of the estimator $\sqrt{\text{Var}(\widehat{\mathcal{O}})}$ (solid), the actual measurement outcome $\text{E}( \widehat{\mathcal{O}} )$ (dotted) and the relative error $\sqrt{\text{Var}(\widehat{\mathcal{O}})}/\text{E}( \widehat{\mathcal{O}} )$ (dashed) for the state $|\psi\rangle \propto \sum_j \psi_j |j\rangle$, with the functions $\psi_j$ indicated in the legend, as a function of power $k$. The GQT (QHP) algorithm is used in the left (right) column. In all cases we consider an $n=6$ qubit circuit, we measure the all zero projector $O = | 0 \rangle \langle 0 |$ and suppose $1000$ shots are used.}\label{fig:PowerErrorComparison}
\end{figure*}

From these two examples, the case of a single qubit system and/or Pauli operators $O$, we suggest that in most cases the QHP converges quicker than the GQT. This is supported by the numerical results shown in Fig.~\ref{fig:PowerErrorComparison}. However, of course, we are not quite comparing like for like here since GQT can be used to implement the transpose operation. 

\subsection{Concatenation can be inefficient}
\label{ap:concatenationcost}

Consider the task of preparing the weighted state $\tau=\rho^{(0)}\odot\rho^{(1)T}$ from the input state $\rho^{(0)}\otimes \rho^{(1)}$. This can be achieved in two ways: (i) use the generalized quantum transpose circuit, but now treating the ancilla register as part of the input, or (ii) first apply the quantum transpose on $\rho^{(1)}$ (by using the GQT circuit with input $\sigma=\ket{+}\!\bra{+}$ and measurement operator $d\times $SWAP) followed by the quantum Hadamard product with $\rho^{(0)}$.  Thus we have two different quantum instruments that implement the same transformation on the inputs: (i) achieves the task directly, and (ii) concatenates two primitives. It is informative to study the sampling complexity of both approaches. We want to estimate the expectation value of some operator $O$ in the weighted state $\tau$ given by $\Tr(\tau O)$.

The variance of the estimator for method (i) is already calculated in Eq.~\eqref{eq:varGQT} and is given by $\Tr(D(\rho^{(0)})O^2)-|\Tr(\tau O)|^2$. The variance of the estimator for method (ii) can be expressed in terms of Eq.~\eqref{eq:Var} if we treat the two concatenated instrument as a single instrument with measurment operator $d(\ket{0}\!\bra{0}\otimes SWAP)$ and is given by $d^2\Tr(D(\rho^{(0)})O^2)-|\Tr(\tau O)|^2$.  The additional factor of $d^2$ drastically favors the first method. This demonstrates that concatenation of primitives can be very costly and it might be advantageous to design instruments that implement a desired transformation directly. 

A similar observation can be made with regards to the LCS algorithm. Given a concatenation scheme to prepare the superposition of $L>2$ states, one can in general find a more efficient way of implementing the same task directly using a single instrument that takes all states as input as described in Section~\ref{ap:allinone}.

\section{Quantum state polynomials realizable without randomization of instruments}\label{ap:RealizableStates}

In Section.~\ref{ap:beyondnormal} we have described how quantum instruments with nonnormal measurement operators can thereby be emulated by physical quantum instruments that have normal measurement operators and can be realized in a quantum computer. This is done by randomizing over quantum instruments. This additional step, however, can increase the sampling complexity, as it adds yet another component to the variance of the estimator.  However, in practice there might be other considerations than minimizing the sampling cost. For instance if $M$ can be expressed as the sum of Pauli operators, the circuits for measuring each term can be much shorter than the one for measuring $M$ directly. In such cases the randomized instrument may be preferred in a noisy implementation. 
Below we classify the class of all 
transformations we can implement with the QSP algorithm using a single instrument shown in Fig.~\ref{fig:QSP} (i.e. without the additional ancilla as in Fig.~\ref{fig:arbitrary_M}).

Suppose $\alpha\in\mathbb{C}^{2\times2}$ contains the coefficients for which that we wish to realize a weighted state. $\alpha$ can be written as $\alpha=H+iS$ where $H=\frac{\alpha+\alpha^\dagger}{2}$ and  $S=\frac{\alpha-\alpha^\dagger}{2i}$ are both Hermitian matrices. Since the Pauli matrices, $\mathcal{P}=\{I,X,Y,Z\}$, span the set of Hermitian matrices, we can write $H=\sum_{P\in \mathcal{P}}h_{P}P$ and $S=\sum_{P\in \mathcal{P}}s_{P}P$. Note that the coefficients, $h_P, s_P$, can be found easily from the entries of $\alpha$.

Let us assume for now that the input states are normalized. We would like to find a density operator $\sigma$ and normal matrix $M$ such that $\alpha=\sigma\odot M^T$.
Since $\sigma$ is a density operator it can be written as $\sigma=I/2+(r/2)\Vec{u}\cdot \Vec{\sigma}$, where $\Vec{\sigma}\equiv(X,Y,Z)$, $\norm{\Vec{u}}=\norm{(u_x,u_y,u_z)}=1$, and $r\in[0,1]$. Similarly, by normality of $M$ we can write $M=(aI + \Vec{v}\boldsymbol{\cdot}\Vec{\sigma}) + i(bI + c\Vec{v}\boldsymbol{\cdot}\Vec{\sigma})$ where $a,b,c\in\RR$ and $\Vec{v}\in\RR^3$. Without loss of generality we let $u_y=0$, so that $\Vec{u}=(\sin\theta, 0, \cos\theta)$.

Comparing this choice of coefficients with the requirement $\alpha=\sigma\odot M^T$, we find:
\begin{align}
    H &= h_I I + h_X X + h_Y Y + h_Z Z = (a+r\cos\theta v_z)I+(r\sin\theta v_x)X - (r\sin\theta v_y)Y + (ar\cos\theta +v_z)Z \label{eq:hermitian}\\
    S &= s_I I + s_X X + s_Y Y + s_Z Z = (b+cr\cos\theta v_z)I+(cr\sin\theta v_x)X - (cr\sin\theta v_y)Y + (br\cos\theta +cv_z)Z \label{eq:skew}
\end{align}

It can easily be shown that choosing $\sigma$ to be pure is more general than choosing a mixed state, so we will assume $r=1$. We also only consider the case where $\alpha$ is not diagonal, since this can be dealt with easily. Note that this condition will necessarily require $\sin\theta$ and $\cos\theta$ to be nonzero.

Equations \eqref{eq:hermitian} and \eqref{eq:skew} represent 8 equations in 7 variables, $a,b,c,v_x,v_y,v_z,$ and $\theta$. By comparing the equations for $h_X, h_Y, s_X$, and $s_Y$, we immediately see that $s_x = c h_x$ and $s_y = c h_y$. These imply that $c=s_x/h_x = s_y/h_y$, which is easily shown to be equivalent to $\abs{\alpha_{01}}=\abs{\alpha_{10}}$. We can satisfy these four equations by setting
\begin{align}
    v_x &= \frac{2h_x}{\sin\theta}, &v_y = \frac{-2h_y}{\sin\theta}.
\end{align}
We now consider two cases.

\textbf{Case 1: $\alpha=e^{i\phi}\Hat{H}$ for some Hermitian $\Hat{H}.$} By Euler's formula, we know that $\alpha=\cos\phi\Hat{H}+i\sin\phi\Hat{H}=H+iS$. Thus, for every $i\in\{I,X,Y,Z\}$ it is implied that $s_i=\tan\phi \cdot h_i$ (in fact, the assumption is equivalent to the existence of a scalar, $k$, such that $s_i=k \cdot h_i$ for every $i$). By setting $b=a\tan\phi$ and $c=\tan\phi$ we can satisfy the $s_I$ and $s_Z$ equations. By simple substitutions in the $h_I$ and $h_Z$ equations, we find that
\begin{equation}
    a = 2h_0-\cos\theta v_z,\;\;v_z=\frac{2h_z-2h_0\cos\theta}{1-\cos^2\theta}.
\end{equation}
We have solutions for $a,b,c,v_x,v_y,$ and $v_z$, and all equations are satisfied, but $\theta$ is still a free parameter. Thus in the case $\alpha=e^{i\phi}\cdot \Hat{H}$ for some Hermitian $\Hat{H}$, we have an infinite choice of ancilla states by setting $\theta$ to be any non-multiple of $\pi$.

\textbf{Note:} the assumption in this case is a generalization of both Hermitian and skew-Hermitian matrices. Suppose $A\in\mathbb{C}^{2\times 2}$ satisfies $A=e^{i\phi}\Hat{H}$. This occurs if and only if $A^\dagger=e^{-i\phi}\Hat{H}= e^{-2i\phi}e^{i\phi}\Hat{H}=e^{-2i\phi}A$. Note that if $\phi=0$ then $A$ is Hermitian, and if $\phi=\pi/2$ then $A$ is skew-Hermitian.

\textbf{Case 2: $\alpha\neq e^{i\phi}\cdot \Hat{H}$ for any Hermitian $\Hat{H}.$}  By solving the equations for $h_I$ and $s_I$ for $a$ and $b$ and substituting these values into the equations for $h_Z$ and $s_Z$, we find that
\begin{equation}
    \cos\theta = R(\alpha)\equiv\frac{s_z - c h_z}{s_0 - c h_0} \label{eq:Ralpha}
\end{equation}
which is well defined since $c=s_0/h_0$ would imply a satisfying assignment such that $c=s_z/h_z$, which contradicts the assumption. Since $\cos\theta$ must be strictly between -1 and 1 ($\alpha$ is not diagonal), if $R(\alpha)\notin(-1,1)$ then we will not be able to satisfy \eqref{eq:Ralpha}, implying that no ancilla states exist for $\alpha=\sigma\odot M^T$. However, in the case $R(\alpha)\in(-1,1)$ we can set $\theta = \pm\arccos[R(\alpha)]$, fixing $\theta$ and satisfying \eqref{eq:Ralpha}.

Since $\theta$ is now fixed, the equations for $h_I$, $s_I$, $h_z$, and $s_Z$ represent a linear system of 4 equations in three variables $a,b$, and $v_z$, which can be written as 
\begin{equation}
    \begin{bmatrix}
        1 & 0 &  \cos\theta \\
        0 & 1 & c \cos\theta \\
        \cos\theta & 0 & 1 \\
        0 & \cos\theta & c
    \end{bmatrix}\cdot
    \begin{bmatrix}
        a \\ b \\ v_z
    \end{bmatrix} = 
    \begin{bmatrix}
        h_0 \\ s_0 \\ h_z \\ s_z
    \end{bmatrix},
\end{equation}
which we will abbreviate as $A\Vec{x}=\Vec{y}$. Note that given $\Vec{y}$ this equation will have a unique solution if and only if $A$ is non-singular, i.e. $\det{(A^T A)}\neq 0$. By direct calculation $\det{(A^T A)} = (1+c^2)(1-\cos^2\theta)^4\neq 0$, thus we will be able to determine satisfying assignments for $a, b,$ and $v_z$ by Gaussian elimination.

For this case we see that exactly two pure ancilla states exist for $\alpha=\sigma\odot M^T$ whenever $\abs{\alpha_{01}}= \abs{\alpha_{10}}$ and $R(\alpha)\in(-1,1)$, and no ancilla state (pure or mixed) exists otherwise.

\section{Hardware implementation}\label{ap:HardwareImp}

\begin{figure*}[t!]
    \centering
    \includegraphics[width=0.98\textwidth]{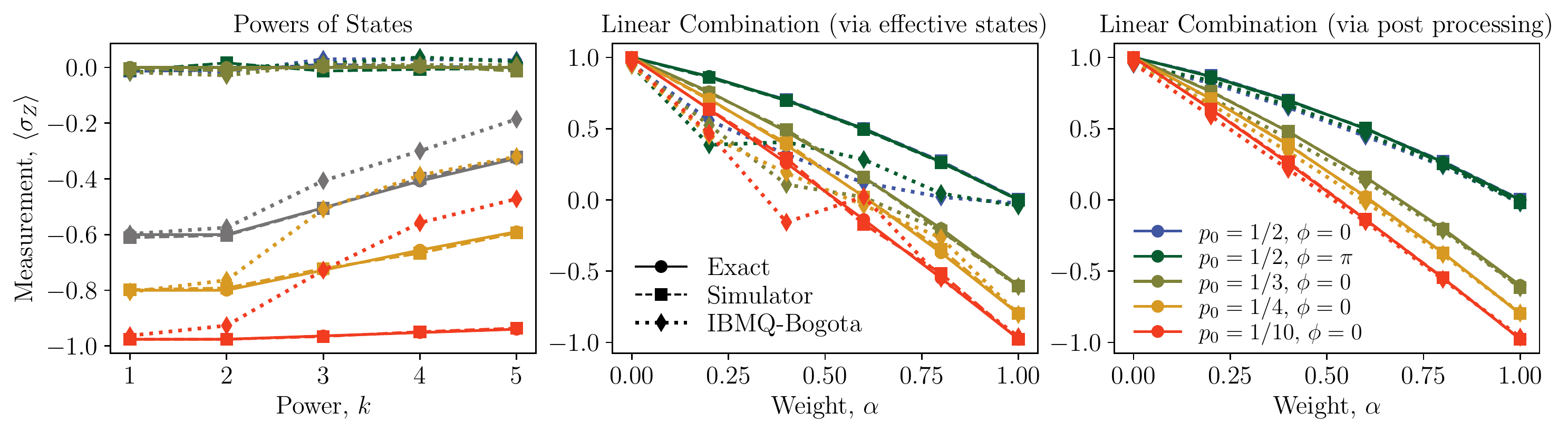}
    \caption{\textbf{Hardware Implementation.} On the left we plot $\bra{\psi(p_0, \phi)^k } \sigma_z \ket{\psi(p_0, \phi)^k}$ where $\ket{\psi(p_0, \phi)} := \sqrt{p_0} \ket{0} + \exp(-i\phi)\sqrt{1  - p_0^2} \ket{1}$ and $\psi^k$ is computed using QHP on a simulator (dashed line) and on IBMQ-Bogota (dotted line). The solid line indicates the exact expectation value computed classically. In the centre (right) plot we plot $\bra{\Psi_\alpha(p_0, \phi) } \sigma_z \ket{\Psi_\alpha(p_0, \phi)}$ where the linear combination of states $\ket{\Psi_\alpha(p_0, \phi)} := \alpha \ket{0} + (1- \alpha) \ket{\psi(p_0, \phi) }$ is computed using the weighted states method (post-processing method). In both the hardware and noiseless simulations $s = 8000$ shots are used. }\label{fig:IBMQ}
\end{figure*}

We implemented our algorithms for performing nonlinear operations on IBMQ-Bogota. In Fig.~\ref{fig:IBMQ} we plot the results of the hardware implementation (dotted) as compared to the ideal output of the algorithm in the absence of hardware noise (dashed) and the ideal output computed classically (solid). The powers of state algorithm (via QHP) works reasonably well for squaring a state, i.e. $k =2$, but performs more poorly for higher powers. We suggest that this is because qubit resets are currently rather noisy. We find that on current hardware the post-processing method for implementing a linear combination of states substantially outperforms the weighted state method. This is perhaps unsurprising given that a controlled swap operation, requiring 18 CNOTs, is required for the weighted states algorithm and the CNOT error on IBMQ-Bogota is of the order $10^{-2}$. More generally, while the hardware results broadly reproduce the expected trends, moderate errors are observed. Thus we expect these algorithms to prove more useful as we approach the fault tolerant era.

\end{document}